# Surface roughness-informed fatigue life prediction of L-PBF Hastelloy X at elevated temperature

Ritam Pal[a], Brandon Kemerling[b], Daniel Ryan[b], Sudhakar Bollapragada[b], Amrita Basak[a,*]

[a] Department of Mechanical Engineering, The Pennsylvania State University, University Park, PA 16802

[b] Solar Turbines Incorporated, San Diego, CA 92101

* Communicating author: aub1526@psu.edu

## Abstract

Additive manufacturing, especially laser powder bed fusion (L-PBF), is widely used for fabricating metal parts with intricate geometries. However, parts produced via L-PBF suffer from varied surface roughness which affects the dynamic or fatigue properties. Accurate prediction of fatigue properties as a function of surface roughness is a critical requirement for qualifying L-PBF parts. In this work, an analytical methodology is put forth to predict the fatigue life of L-PBF components having heterogeneous surface roughness. Thirty-six Hastelloy X specimens are printed using L-PBF followed by industry-standard heat treatment procedures. Half of these specimens are built with as-printed gauge sections and the other half is printed as cylinders from which fatigue specimens are extracted via machining. Specimens are printed in a vertical orientation and an orientation of 30° from the vertical axis. The surface roughness of the specimens is measured using computed tomography and parameters such as the maximum valley depth are used to build an extreme value distribution. Fatigue testing is conducted at an isothermal condition of 500 °F. It is observed that the rough specimens fail much earlier compared to the machined specimens due to the deep valleys present on the surfaces of the former ones. The valleys act as notches leading to high strain localization. Following this observation, a functional relationship is formulated analytically that considers surface valleys as notches and correlates the strain localization around those notches with fatigue life, using the Coffin-Manson-Basquin and Ramberg-Osgood equation. Using the extreme value distribution of the maximum valley depth, the functional relationship is generated with the average of the distribution. The mean life curve from the functional relationship shows a maximum difference of 2% from the experimental mean fatigue life observations for vertically built rough specimens and 10% for 30°-built rough specimens. In conclusion, the proposed analytical model successfully predicts the fatigue life of L-PBF specimens at an elevated temperature undergoing different strain loadings.

Keywords: Laser powder bed fusion; Hastelloy X; high-temperature low cycle fatigue; surface roughness; Extreme value statistics

## 1. Introduction

Additive manufacturing (AM) techniques have paved efficient pathways to fabricate metal parts with complex geometries [1], [2]. For metal AM, techniques involve directed energy deposition [3], powder bed fusion [4], and binder jetting [5]. Among all metal AM techniques, powder bed fusion specifically using lasers, provides unique advantages over other techniques such as minimal material wastage, efficiency in fabricating parts with intricate geometries, and minimal support structure usage [6]. In laser powder bed fusion (L-PBF), powder is spread on a platform which is then consolidated by a laser beam to a desired shape [7]. The L-PBF process witnesses rapid heating and cooling rates that lead to non-equilibrium microstructures that dictate the properties of the fabricated components [8], [9], [10]. Moreover, the L-PBF technique is susceptible to defects such as lack-of-fusion and gas pores [11], [12]. Therefore, a thorough optimization of the process parameters is often conducted in producing defect-free parts with desired mechanical properties [13], [14].

L-PBF technique is widely used to fabricate components using many metallic alloys such as Inconel 718 [15], [16], [17], SS316L [18], [19], [20], AlSi10Mg [21], [22], [23], and Hastelloy X (HX) [24], [25], [26]. The weldability of HX, attributed to low Ti+Al content, makes the material suitable for fabricating complex parts via L-PBF [27], [28]. HX is a nickel-based superalloy with a high oxidation resistance at elevated temperatures [29], [30]. Therefore, HX finds its application in high-temperature operations such as fuel injectors in gas turbines [31], [32], [33] which often involve dynamic loading conditions [34], [35]. In the open literature, the fatigue behavior of wrought HX at elevated temperatures has been reported extensively. For example, Yoon et al. studied the low cycle fatigue life variation of HX at temperatures of 760 and 870 °C which are typical in gas turbine operation [36]. A trapezoidal total strain waveform was applied to the specimens to include tensile and compressive hold times of 1s each. The results showed the fatigue life was lowest at 760 °C because elongation was minimal at that temperature and elongation started to increase from 760 °C to 870 °C. Similar work was reported by Lu et al., where they conducted isothermal fatigue investigations of HX at 816 °C and 927 °C [37]. During fatigue investigation, they introduced various hold times at the peak load to investigate the hold time effect on fatigue crack growth. It was observed that the crack grew faster at higher temperatures, and the crack growth rate increased with the increase in hold time. Apart from the investigation of hold time effects, studies have been conducted on the effect of temperature on the low cycle fatigue behavior of HX. Hong et al. conducted an isothermal low cycle fatigue life investigation of wrought HX at the temperature range of 650-870 °C with various strain ranges [38]. The results showed that cyclic stability was achieved at 870 °C where the fatigue crack initiated at the oxidized grain boundary rather than grain interior, unlike the other temperature conditions where cyclic strain hardening was observed. Similar works have also been reported regarding the effect of temperature on the low cycle fatigue behavior of wrought HX [39], [40]. However, limited work has been conducted on fatigue investigation of L-PBF HX at elevated temperatures [37], [41], [42]. It was observed that fatigue life decreased with the increase in temperature and intergranular crack growth became predominant at elevated temperatures.

One of the primary setbacks of L-PBF components lies in the surface roughness of the components [43], [44]. A significant amount of work has been performed to understand the reasons behind the high surface roughness of components [45], [46]. For example, Tian et al. investigated the effect of fabrication-related process parameters on the surface roughness of HX parts [47]. The surface roughness was found to increase with an increase in hatch spacing because a large hatch spacing caused a small overlap between two consecutive melt pools. The build orientation of the specimens also affected the surface roughness as the heat conduction medium was different for up-skin and down-skin surfaces. For the up-skin surface, conduction was primarily through the previously built solidified layer while the heat conduction was primarily through loose powder particles for the down-skin surface. Laser power and scanning speed also caused the surface roughness of the specimens to change. With the increase in laser scanning speed, balling was observed, which deteriorated the surface quality. A similar study was conducted by Esmaeilizadeh et al. to understand the effect of scanning speed on the surface roughness of L-PBF HX specimens [48]. It was found that a high laser scanning speed resulted in a deeper roughness valley than a low laser scanning speed. The deep roughness valley led to high stress concentration around those regions, which compromised the fatigue properties.

The impact of surface roughness on fatigue properties of L-PBF AM components has been reported in the open literature. For example, Maleki et al. investigated the effect of surface morphology on fatigue properties of L-PBF AlSi10Mg specimens [49]. The specimens with printed surface roughness showed poor surface quality due to the presence of spatter, partially melted, and unmelted powder particles. The poor surface quality deteriorated the fatigue strength of the specimens compared to the shot-peened specimens. Furthermore, Plessis et al. performed computed tomography of L-PBF AlSi10Mg specimens to identify the dominant notches on the rough surfaces that would lead to fatigue crack initiation [50]. It was reported that the notches deeper than 50 µm would act as critical notches during fatigue cycling. Similar investigations were conducted with L-PBF Ti6Al4V specimens by Emuakapor et al [51]. The work showed that the rough specimens demonstrated poor fatigue performance compared to the machined specimens because the former ones had poor surface quality. The impact of surface roughness is more pronounced in low cycle fatigue life investigation as reported in the literature [52], [53]. A limited amount of work has been conducted to understand the impact of surface roughness on the low cycle fatigue properties of L-PBF HX specimens.

As surface roughness affects the fatigue properties of L-PBF components, the prediction of fatigue properties of those components is important yet challenging. Researchers have explored different methodologies to predict fatigue properties of rough L-PBF specimens. Primarily, simulation and machine learning strategies have been combined in the open literature to create a predictive framework. For example, Elangeswaran et al. trained a Gaussian machine-learning model with an experimental dataset to predict the fatigue life of L-PBF SS316L specimens [54]. The machine learning model was augmented with finite element simulations to predict the high cycle fatigue life of rough specimens. However, the work involved extensive experimentation and the finite element simulations required an appreciable amount of time and computational budget. The method resulted in a maximum difference of 50% from the experimental observations. Maleki et al. created a deep neural network model with experimental results to predict the high cycle fatigue life of rough L-PBF AlSi10Mg specimens [55]. The model revealed that the surface conditions significantly impacted the fatigue lives of the specimens. For L-PBF Ti6Al4V specimens, Macallister and Becker built a damage-tolerant model with a defect population from computed tomography as input [56]. The model was created to predict the high cycle fatigue life of rough specimens and the results showed that the fatigue life deteriorated when the defect population was close to the boundary of the specimens. Moreover, the model produced a maximum difference of 40% from the fatigue lives of as-fabricated specimens. All the previous studies on predictive models were focused on the high cycle fatigue life of L-PBF specimens and they resulted in a significant difference from the experimental observations. Furthermore, they did not involve fatigue life prediction of rough specimens at an elevated temperature.

In the current work, an analytical methodology is proposed that correlates surface roughness, operating temperature, and fatigue life of L-PBF HX specimens. The specimens are investigated at an isothermal condition of 260 °C (500 °F) that corresponds to the operating conditions of typical land-based gas turbine engines' fuel injectors. The experimental investigations are conducted with specimens built at two different orientations. Each set of orientations contains two different sets of machined and rough specimens. Therefore, in total, four different sets of specimens are investigated at three different strain ranges. The surface roughness of the rough specimens is evaluated using computed tomography and, thereafter, an extreme value distribution is used to obtain the surface roughness parameters. From the fatigue investigation dataset, initiation life is evaluated for all specimens using a compliance methodology, and the evaluation is compared with fractography results. Finally, an equation is derived to predict the fatigue life using the surface roughness parameter, strain ranges, and material properties as functions of temperature such as Coffin-Manson-Basquin parameters and Ramberg-Osgood parameters.

## 2. Material and methodology
### 2.1.1 Specimen fabrication

All specimens are fabricated on a carbon steel buildplate via L-PBF using a Concept Laser M2 system at Solar Turbines Incorporated (San Diego, USA), using gas-atomized HX powder with a nominal composition of Cr 20.5-23%, Fe 17-20%, Mo 8-10%, and Ni (balance). The nominal powder size distribution is 15 – 45 µm as described in Figure 1. Figure 1(a) shows a scanning electron microscope

(SEM) image of the powder feedstock and Figure 1(b) presents the powder particle size distribution showing a skewness around particle sizes of 1 to 30 µm. A volumetric energy density of 70 J/mm³ is used to fabricate the specimens at a layer thickness of 35 µm. The specimens are removed from the build plate using wire-electric discharge machining technique. All specimens are subjected to a hot isostatic press (HIP) cycle using typical settings for industrial applications. Two nickel braze thermal exposures are also applied to all specimens following HIP. and then machined for characterization and testing.

The fatigue specimen design conforms to the ASTM E606 standard [57] with the ratio of gauge diameter to gauge length as 0.3, shown in Figure 2(a). Four different specimen types, shown in Figure 2(b), are fabricated. They are (1) ASTM E606 specimens built in the vertical orientation (2) ASTM E606 specimens built 30° from the vertical orientation, (3) vertically built cylindrical specimens, and (4) cylindrical specimens built 30° from the vertical orientation. While only grip sections of the ASTM E606 specimens are machined to fit the fatigue testing apparatus adequately, the cylinders are fully machined to extract fatigue specimens. Therefore, machined specimens are labeled as '*machined*', while specimens retaining their gauge section roughness are referred to as '*rough*.' The difference between *machined* and *rough* gage sections is depicted in Figure 2(c).

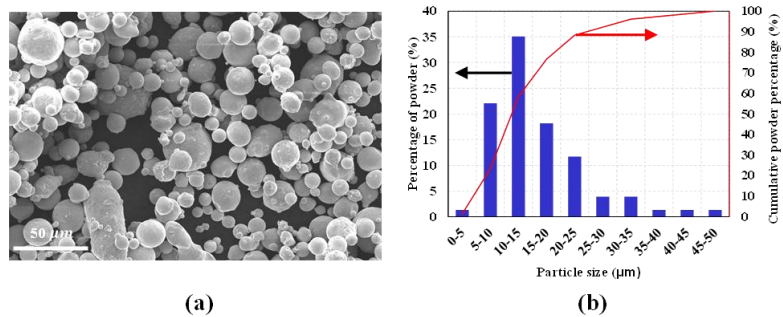

**Figure 1. (a) SEM image of gas-atomized HX powder particles. (b) Powder particle size distribution with nominal size diameter of 15-30 µm.**

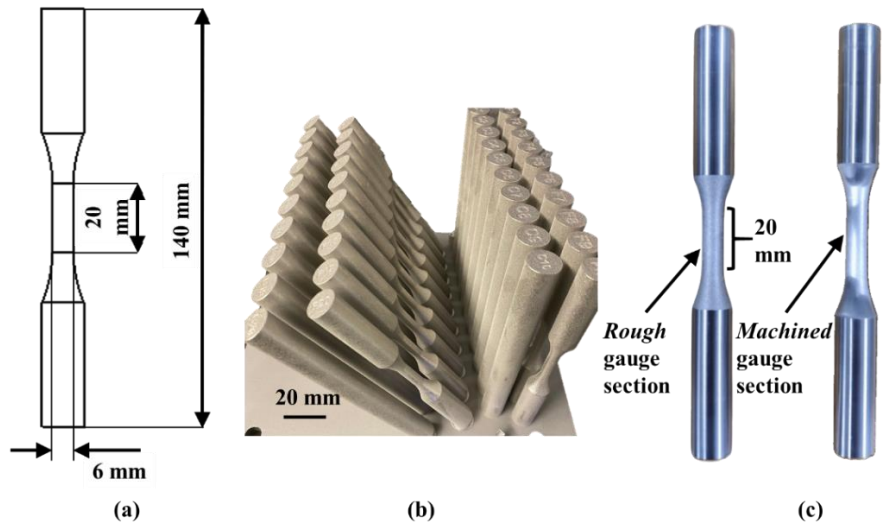

**Figure 2: (a) ASTM E606 design of fatigue specimens. (b) L-PBF fabricated specimens on the build plate showing the four different sets of specimens. (c) Two representative specimens showcasing rough and machined gauge sections.**

## 2.2 Experimental investigation

The experimental investigation encompasses surface roughness evaluation using computed tomography (CT), thermomechanical fatigue life investigation in MTS equipment, and initiation life evaluation using compliance technique and fractography analysis.

### 2.2.1 Surface roughness evaluation from X-ray CT

X-ray micro-CT is performed in a GE v|tome|x system located at Penn State. The equipment can carry out CT imaging at a sub-nanometer resolution. In the current work, the specimens are imaged with 200 kV voltage and 75 µA current, resulting in a 7.5 µm resolution. During the CT imaging, three projections are recorded for each specimen at each acquisition angle utilizing the stepscan technique. The projections are reconstructed in the system's software by employing the filtered back projection algorithm and the reconstruction results in a volume file. The volume file is imported in ImageJ [58] to export the image stack as tagged interchangeable file format files that are compatible with Avizo 3D 2021 software [59]. In Avizo, 3D reconstruction is performed with the image stack after the application of a median filter to mitigate noises. The reconstruction also consists of a thresholding operation where the software separates pores from the solid body.

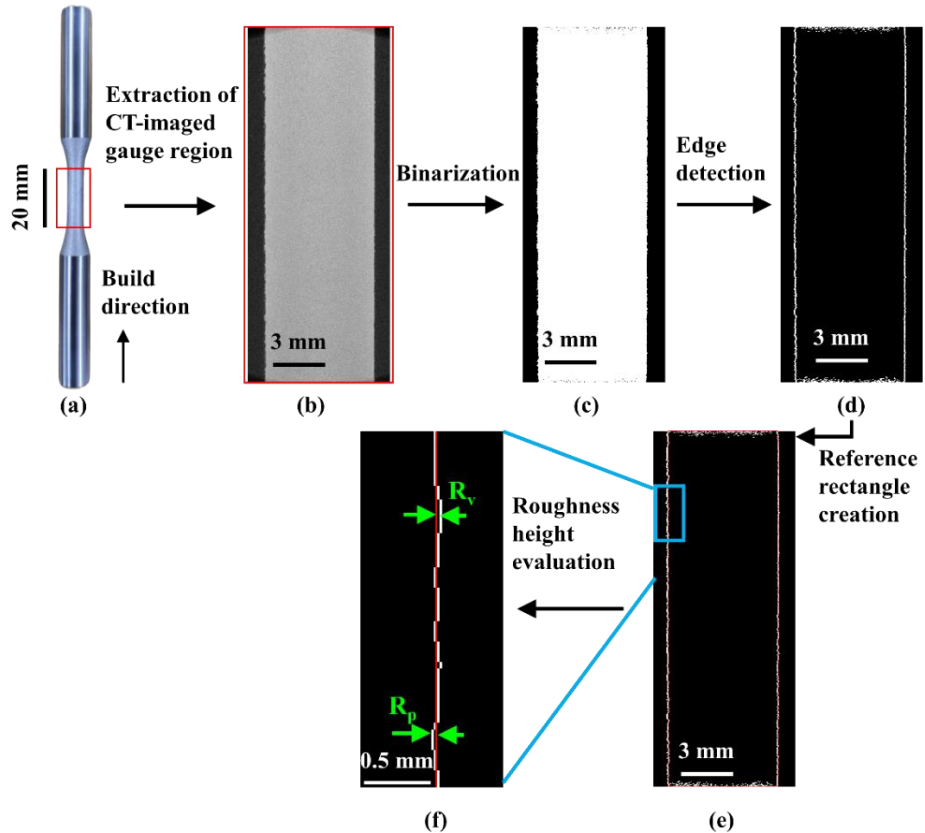

**Figure 3. Schematic representation of the processes involved in the extraction of surface roughness parameters from longitudinal cross-sectional CT images of a representative specimen shown in (a). The extraction of the longitudinal cross-section in (b) is followed by binarization in (c) and edge detection of the image in (d). An ideal rectangle of width 6 mm, corresponding to the specimen diameter, is fitted to the edges, and the roughness profile is extracted from the ideal rectangle and the rough edges in (e). (f) Nomenclature showing peak height and valley depth calculation from (e).**

The roughness of the specimens is evaluated from the stack of image files using two different methodologies. The first method includes roughness evaluation from longitudinal cross-sectional images of the specimens as shown in Figure 3. The gage section corresponds to the area of interest in CT imaging, presented in Figure 3(a). From the image stack, a cross-sectional image is selected that represents the entire diameter of the specimen, presented in Figure 3(b). The image is imported in MATLAB to perform binarization followed by edge detection, shown in Figures. 3(c) and (d), respectively. The reference line for roughness evaluation is chosen in a manner that the distance between those reference lines matches the intended gauge diameter of 6 mm. The Canny edge detection algorithm is employed to detect the rough surface of the specimen or in other words, the rough edges of the 2-dimensional images, presented in Figure 3(d). The pixel position of the rough edges serves as a parameter to quantify the surface roughness of the specimens. The distance between those edge pixels and the reference line pixel positions is used to evaluate the roughness profile, shown in Figure 3(e). The roughness profile shows typical peaks and valleys. From this profile, the peak height ($R_p$) is calculated from the edge pixels which are on the left side of the reference line while the valley depth ($R_v$) is calculated from the edge pixels which are on the right side of the reference line, shown in Figure 3(f). Using these $R_p$ and $R_v$ values, the root-mean-squared (RMS) roughness of the specimens is calculated [60]. Furthermore, maximum valley depths ($R_{v,max}$) are obtained for each specimen. The evaluation of $R_{v,max}$ is discussed in detail in the following sections.

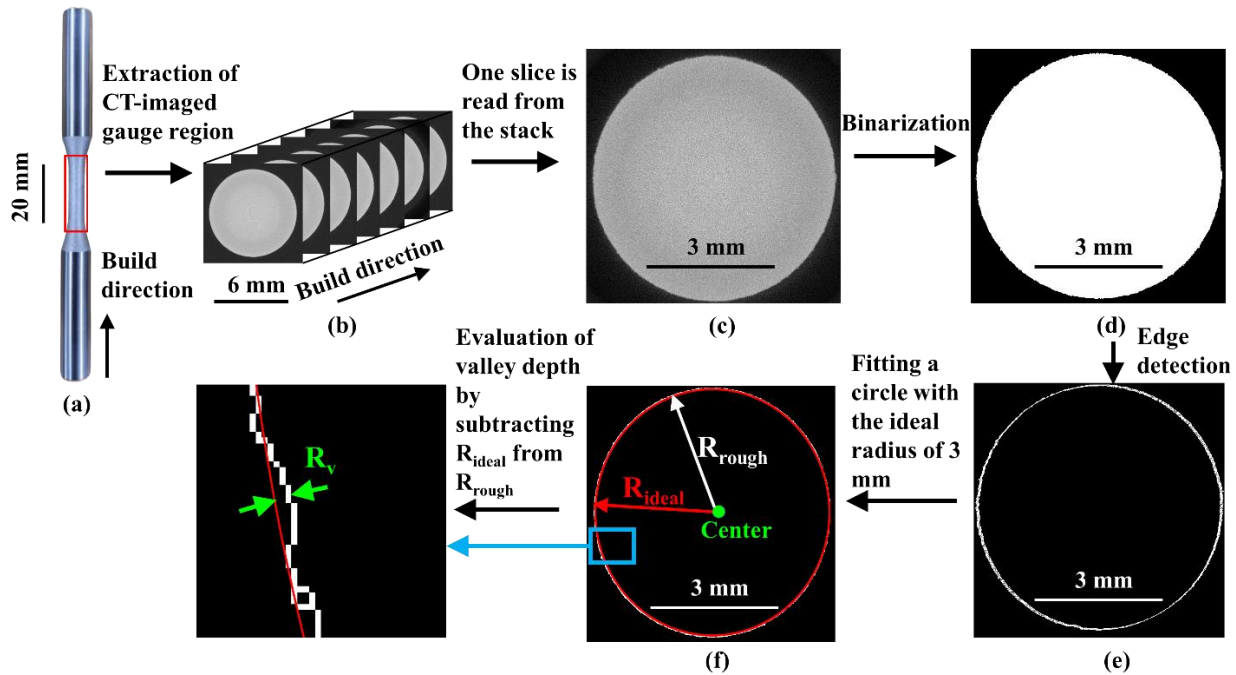

**Figure 4. Schematic representation of the processes involved in the extraction of surface roughness parameters from transverse cross-sectional CT images of a representative specimen shown in (a). (b) Stack of transverse cross-sectional CT images. (c) One of the images is imported in MATLAB. The transverse cross-section image extraction is followed by (d) binarization and (e) edge detection. (f) An ideal circle of diameter 6 mm, corresponding to the specimen diameter, is fitted to the edges, and $R_v$ is extracted from the ideal circle and the rough edges.**

The second method of surface roughness evaluation involves transverse 2D cross-sectional CT images of the specimens. All the transverse cross-sectional images that belong to the gauge region (Figure 4(a)) are imported in MATLAB. The images from the stack (Figure 4(b)) are imported in MATLAB (Figure 4(c)), where they undergo binarization and Canny edge detection, which isolates the rough edge from the solid volume, shown in Figures 4(d) and (e), respectively. To evaluate the surface roughness profile, a reference circle is created about the center point of the image such that the diameter of the reference circle matches the design intent gauge diameter of 6 mm, shown in Figure 4(f). The surface roughness parameters are obtained using the distance between the pixel positions of the rough edge and the reference circle's pixel positions. From each cross-sectional image, RMS roughness and $R_{v,max}$ are evaluated. For an entire specimen, RMS roughness is obtained by evaluating the average of the RMS roughness of all the cross-sectional images of that particular specimen. On the other hand, the maximum valley depth of each image is used to create a Gumbel distribution for the entire specimen. Gumbel distribution refers to Type-I generalized extreme value distribution where the maximum or the extreme values of a particular property are used to obtain the average of the distribution [61]. In this case, the maximum valley depths are used to build the distribution, and the average of the distribution is considered the maximum valley depth of the specimen. The maximum valley depth of each specimen is further employed in formulating a correlation between the surface roughness and fatigue life of the specimens. It is discussed in detail in section 4.4.

### 2.2.2 Total fatigue life evaluation

The test frame is an MTS® 370.10 thermomechanical fatigue test system, which can be configured for use with standard servo-valves. Standard MTS® 646 hydraulic collet grips (load capacity: 100 kN) are available for holding test specimens with round grip sections. The test controller is an MTS® FlexTest 40 digital controller. The test control software is MTS TestSuite MPE 3.6.6 and is hosted on a PC with a 3.3 GHz processor, and 16 GB RAM running the Windows 10 operating system. The control software has toolboxes enabled for different command feedback compensation schemes. An induction coil is used to impose temperature on the specimens and the temperature is controlled by a generator manufactured by TrumpF Inc. Temperature is accurately calibrated by a feedback loop that consists of 2 k-type thermocouples (manufactured by OMEGA). A high-temperature knife-edge extensometer (Model MTS 632.53F-14) is used to measure strain.

The low-cycle fatigue life investigation consists of 35 specimens. An experiment matrix with specimen IDs is presented in Table 1. The specimen ID is derived considering the build direction (i.e., V for vertical and 30 for 30° from vertical) and gauge condition (i.e., M for machined and R for rough). Four different sets of specimens are tested: nine VR, eight VM, nine 30R, and nine 30M. Three different total strain ranges: 0.7%, 1%, and 1.5%, are employed at a loading (R) ratio of -1 to carry out the fatigue tests for each set of specimens. While the original plan had nine VM specimens, one VM specimen was lost due to experimental issues. The experiment encountered unexpected noises and the specimen failed without producing any meaningful results. Hence, within the test matrix, while every strain condition has 3 repeats, the 0.7% strain range has 2 repeats because it produces very consistent fatigue results. The strain values are imposed on the specimens as a triangular waveform with a frequency of 1 Hz for 0.7% and 1% strain ranges, and 0.33 Hz for 1.5% strain range. These frequencies ensure that the strain rate is between 0.01 to 0.02 s$^{-1}$ to avoid creep failures at a low strain range [62]. At each strain range, three specimens from the same set are used to investigate the repeatability of the experiments. All the experiments are carried out at an isothermal condition of 500 °F because this temperature corresponds to the operating temperature of a fuel injector in a gas turbine [63]. In the experiments, fatigue failure of a specimen is considered when complete separation has taken place.

**Table 1. Fatigue investigation matrix of all L-PBF HX specimens. The specimen ID is derived considering the build direction (i.e., V for vertical and 30 for 30° from Vertical) and gauge condition (i.e., M for machined and R for rough).**

| Specimen ID | Test temp (°F) | R ratio | Specimen print angle | Gauge section surface condition | Total Strain Range (%) |
|---|---|---|---|---|---|
| VR1 | 500 | -1 | Vertical | As-built L-PBF ASTM E606 specimens that are heat treated, referred to as L-PBF roughness or '*rough*' | 1.5 |
| VR2 | | | | | 1.5 |
| VR3 | | | | | 1.5 |
| VR4 | | | | | 1 |
| VR5 | | | | | 1 |
| VR6 | | | | | 1 |
| VR7 | | | | | 0.7 |
| VR8 | | | | | 0.7 |
| VR9 | | | | | 0.7 |
| VM1 | | | | As-built L-PBF cylindrical specimens that are heat treated and machined, referred to as '*machined*' | 1.5 |
| VM2 | | | | | 1.5 |
| VM3 | | | | | 1.5 |
| VM4 | | | | | 1 |
| VM5 | | | | | 1 |
| VM6 | | | | | 1 |
| VM7 | | | | | 0.7 |
| VM8 | | | | | 0.7 |
| 30R1 | | | 30° from vertical | *rough* | 1.5 |
| 30R2 | | | | | 1.5 |
| 30R3 | | | | | 1.5 |
| 30R4 | | | | | 1 |
| 30R5 | | | | | 1 |
| 30R6 | | | | | 1 |
| 30R7 | | | | | 0.7 |
| 30R8 | | | | | 0.7 |
| 30R9 | | | | | 0.7 |
| 30M1 | | | | *machined* | 1.5 |
| 30M2 | | | | | 1.5 |
| 30M3 | | | | | 1.5 |
| 30M4 | | | | | 1 |
| 30M5 | | | | | 1 |
| 30M6 | | | | | 1 |
| 30M7 | | | | | 0.7 |
| 30M8 | | | | | 0.7 |
| 30M9 | | | | | 0.7 |

### 2.2.3 Initiation life evaluation
#### 2.2.3.1 Compliance method

The initiation life of the specimens is evaluated by employing a compliance technique. In this methodology, raw load (F)-displacement (D) data is converted to compliance (C) values as shown in equation (1). For each loading cycle, there is a compliance value associated with the peak load and peak displacement values. Here, peak load is considered because a crack always opens at the peak load of the loading cycle [64]. At first, compliance values are plotted against displacement values, and the slope is evaluated from the plot. The compliance-displacement slope {(C2-C1)/(D2-D1)} is plotted against loading cycles to observe the marked increase in compliance slope. The demarcated change in compliance slope is considered the crack initiation cycle in a specimen. The compliance slope increases due to the load drop and displacement increase occurring across the specimen during crack initiation which is amplified with the evaluation of the compliance slope [64]. For representation, the compliance difference values are plotted against cycles as discussed in section 3.1. A detailed discussion of the compliance variation phenomenon is described in section 4.4.

$$C = D/F \tag{1}$$

#### 2.2.3.2 Fractography method

After the specimens' fatigue investigation, fractography analysis is conducted using an SEM. All the micrography images are acquired with a scanning voltage of 25 kV and a current of 65 nA using a Q250 analytical scanning electron microscope (Thermo Scientific, USA). The primary objective of fractography analysis is to evaluate the initiation life of the specimens by observing the initiation zone and the propagation or beach marks [65]. For initiation life evaluation, different parameters are obtained from the fractured surface such as the initial crack length and the ratio of initial and final crack length. These parameters are utilized in the following equations [66] and the resultant initiation life is used to validate the initiation life obtained from compliance methodology.

$$(K_0)_{max} = F_0 \sigma_{max} \sqrt{\pi a_0} \tag{2}$$

$$(K_0)_{min} = F_0 \sigma_{min} \sqrt{\pi a_0} \tag{3}$$

$$F_0 = g\left(0.952 + 2.02\lambda + 0.37\left(1 - \sin\left(\frac{\pi\lambda}{2}\right)\right)^3\right) \tag{4}$$

$$g = 0.92 \left(\frac{2}{\pi}\right) \sqrt{\frac{\tan\left(\frac{\pi\lambda}{2}\right)}{\frac{\pi\lambda}{2}}} \left(\frac{1}{\cos\left(\frac{\pi\lambda}{2}\right)}\right) \tag{5}$$

$$N_f - N_i = \left(\frac{2a_0}{(m-2)C(\Delta K_0)^m}\right)\left[1 - \left(\frac{a_0}{a_f}\right)^{\frac{m}{2}-1}\right] \tag{6}$$

Here, $\sigma_{max}$ and $\sigma_{min}$ are maximum and minimum stress of stable cycle, respectively, $(K_0)_{max}$ and $(K_0)_{min}$ are the maximum and minimum stress intensities at the crack initiation cycle, respectively, $F_0$ is the magnification factor based on crack lengths, g is the geometry factor assuming the crack has a circular arc front, $a_0$ is the initial crack length, $a_f$ is the final crack length, $\lambda$ is the ratio of initial crack length and gauge diameter, C is the Paris law coefficient, m is the Paris law exponent, $N_f$ is the final fatigue life, and $N_i$ is the initiation life. The values for C and m are $1.8^{-12}$ and 5, respectively, and obtained from the literature [67].

## 3. Results

This section summarizes the results obtained from CT imaging regarding the surface roughness of the *rough* specimens followed by their fatigue lives. The initiation life of the specimens is also put forth using compliance technique and fractography analysis.

### 3.1 Surface roughness evaluation

The surface roughness of all specimens is analyzed following the method described in section 2.2.1, where longitudinal and transverse cross-sectional CT images are processed in MATLAB to evaluate surface roughness parameters. The parameters include RMS roughness and maximum valley depth for each specimen. The results from longitudinal cross-sections are presented in Figure 5(a). The 30R specimens show higher roughness on their down-skin surface than their up-skin surface. The RMS roughness on the down-skin surface (denoted by D) ranges from 22 $\mu m$ to 35 $\mu m$. On the contrary, up-skin surfaces (denoted by U) showcase RMS roughness from 9 $\mu m$ to 17 $\mu m$. The VR specimens show surface roughness homogeneity across the entire gauge section and the RMS roughness is similar to the up-skin surfaces of 30R specimens. The surface roughness of the VM and 30M specimens is not evaluated from CT imaging as the roughness variation in those specimens is much less than the CT scan resolution of 10 $\mu m$ which can be observed from Figure 5(b). Furthermore, Figures 5(c) and (d) illustrate the surface roughness of two representative VR and 30R specimens, confirming the observations in Figure 5(a).

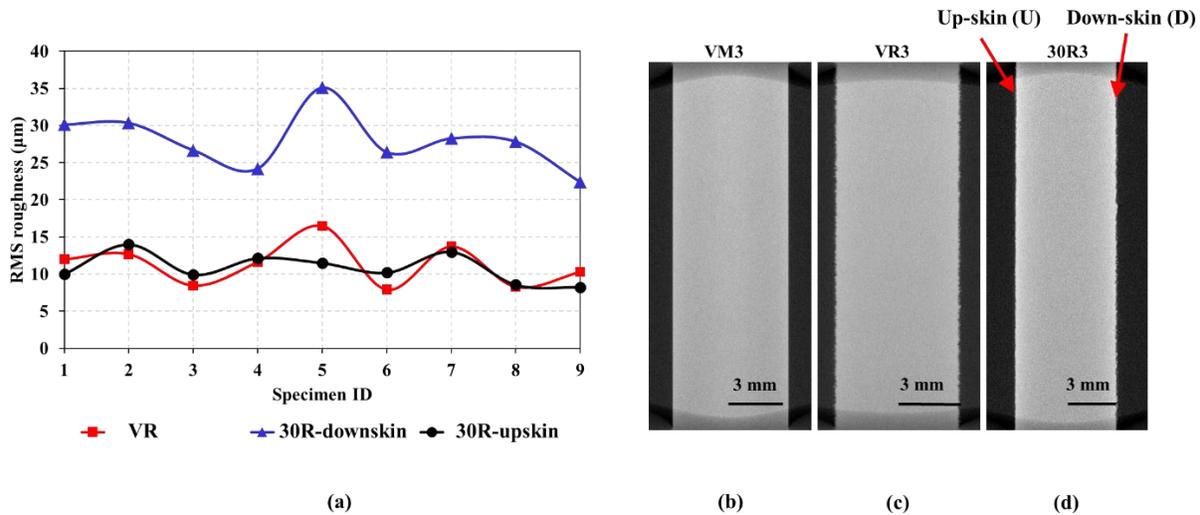

**Figure 5. (a) Comparison of RMS roughness between vertical and 30° built specimens. Here, U denotes up-skin surfaces and D denotes down-skin surfaces. Longitudinal cross-sectional images of (b) VM6, (c) VR6, and (d) 30R6 specimens. The 30° built specimens consist of up-skin and down-skin surfaces, and the down-skin surface shows a significant increase in surface roughness.**

### 3.2 Fatigue life evaluation

The four sets of specimens are investigated at different strain ranges at an elevated temperature of 500 °F. At each strain range, three fatigue tests are conducted to observe the repeatability. A bar chart is presented in Figure 6(a), which depicts the variation of fatigue life of different sets of specimens tested at different strain ranges. The VM specimens show the maximum fatigue life across all strain ranges whereas the 30R specimens yield the minimum fatigue life at all strain ranges. The VR specimens' fatigue life is almost half of that of VM specimens, which implies that the surface roughness significantly impacts fatigue life. The average fatigue lives of 30R specimens are 1,629, 4,628, and 9,882, respectively for 1.5%, 1%, and 0.7% strain ranges which are within 14% of the average fatigue lives of VR specimens for the same strain ranges. The difference of 14% can be attributed to the higher roughness of 30R specimens than the VR specimens but the difference is less compared to the 50% difference between VR and VM specimens. Similar observations can be made for VM and 30M specimens which show that the average fatigue lives of 30M specimens are within 8% of the VM specimens' average fatigue lives for 1 % and 1.5% strain ranges. This suggests that at high strain ranges, build orientation does not have a significant impact on the fatigue lives of machined specimens. At high strain ranges, transgranular crack initiation takes place which is independent of the grain structure and grain orientation, as reported in the literature for HX [38]. However, for the 0.7% strain range, 30M specimens show a difference of 18% from VM specimens' average fatigue live value. It implies that at lower strain ranges, build orientation starts to influence the fatigue life of machined specimens. With the decrease in strain range, intergranular crack initiation occurs which depends on the grain structure orientation and hence, build direction [38]. From the observations, it can be highlighted that surface conditions have more impact on the low cycle fatigue lives than the build orientation. Further microstructural characterization needs to be conducted in the future to shed more insight into transgranular and intergranular crack growth.

From the repeatability investigation of each specimen set at a particular strain range, average fatigue life values are calculated, and the values are plotted against the strain ranges. The plot in Figure 6(b) shows the increase in fatigue life with the decrease in strain amplitude. Moreover, the figure presents the scatter of fatigue life for each test condition with a maximum scatter of 5% for the 30R specimen set at a 0.7% strain range. The trend of fatigue life variation with strain range is similar for all specimen categories. The fatigue life variation curves of VM and 30M specimens almost overlap each other, which suggests that they follow the Coffin-Manson-Basquin relation of low cycle fatigue life. However, the *rough* specimens show significant deviation from the *machined* specimens, irrespective of the same fabrication and experimental conditions. In addition, 30R specimens deviate from the results of VR specimens. This suggests that surface roughness plays a crucial role in low cycle fatigue investigations of *rough* specimens.

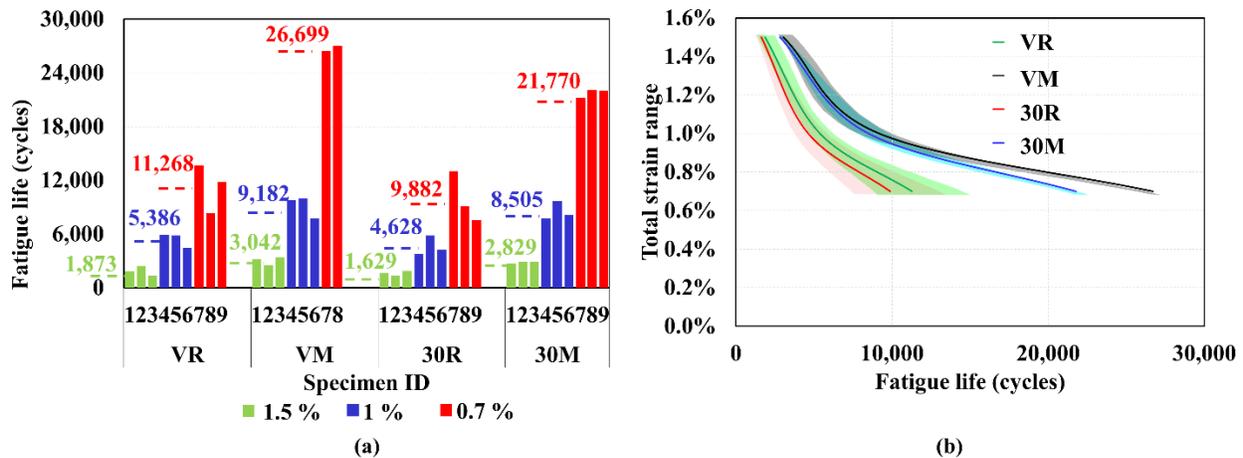

**Figure 6. (a) Variation of fatigue life for different sets of specimens and (b) mean fatigue life curves of different sets of specimens with shaded regions of each curve showing the variance of a particular specimen set. In (a), the average fatigue life of specimens corresponding to a strain range set is shown by dashed lines.**

## 3.3 Initiation life evaluation

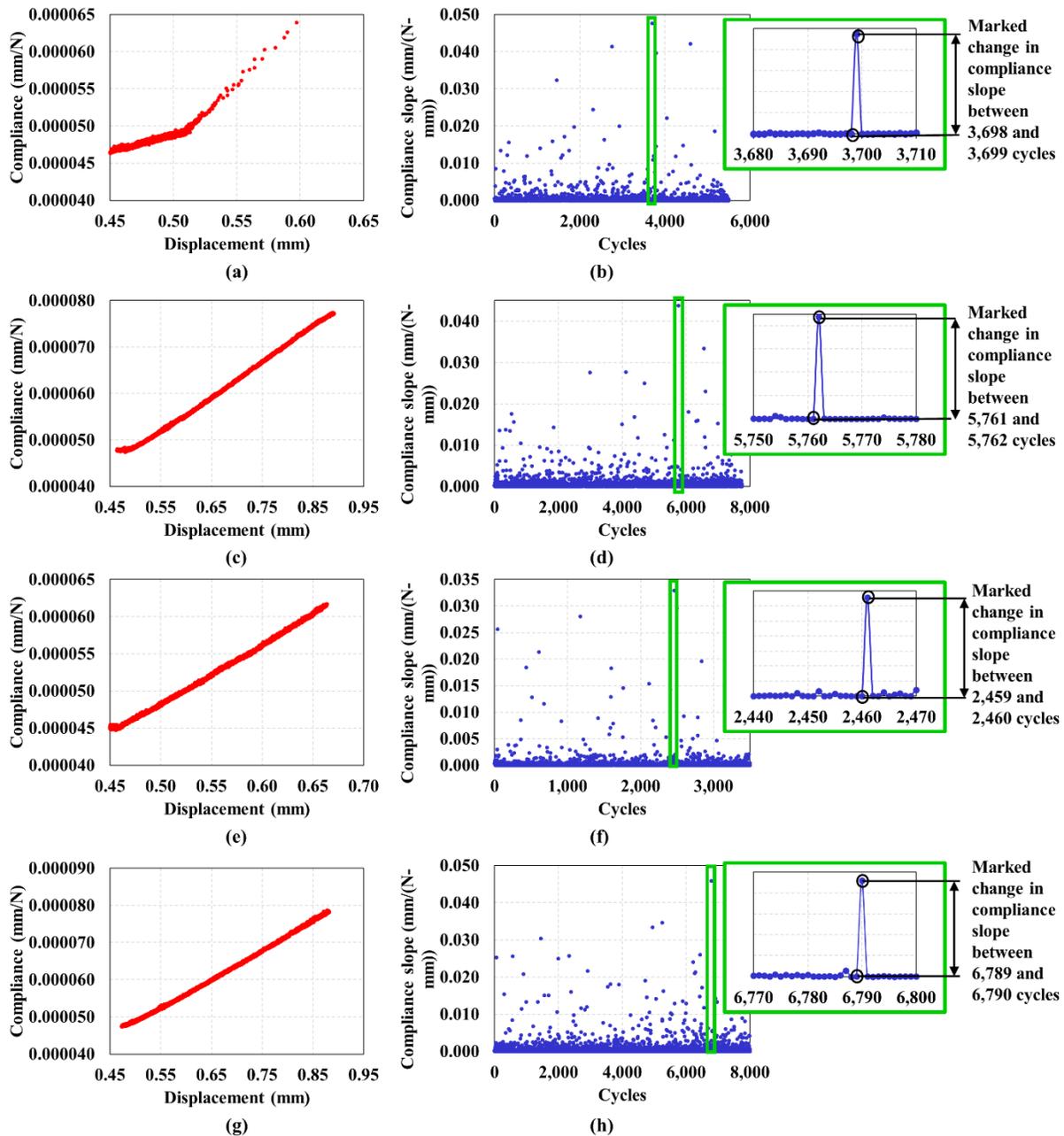

**Figure 7.** Compliance-displacement curves of (a) VR5, (c) VM5, (e) 30R4, and (g) 30M4. Variation of compliance-displacement slope with loading cycles is shown for (b) VR5, (d) VM5, (f) 30R4, and (h) 30M4. All the specimens are tested at a 1% total strain range. The compliance-displacement slope variation consists of a marked change in slope between two consecutive cycles, indicating crack initiation.

From the fatigue investigations, load-displacement data is collected, and compliance is evaluated from the load-displacement data. The variation of compliance with displacement is shown in Figure 7(a) for a representative specimen, VR5. It can be observed that after a certain time, the compliance increases with the displacement, and there is a change in the compliance-displacement slope. This slope is plotted against the number of cycles to locate the cycle at which there is a notable change, shown in Figure 7(b). The demarcated change in compliance marks the crack initiation cycle in the specimen. The compliance-displacement slope increases at crack initiation because a load drop and a displacement increase are associated with crack initiation which increases the specimen's compliance during that cycle. A similar procedure is repeated for all the sets of specimens to evaluate their initiation life at different experimental conditions. The compliance-displacement and compliance slope-cycles plots for a few representative specimens, VM5, 30R4, and 30M4 are shown in Figures 7(c), (d), (e), (f), (g), and (h), respectively. All the specimens show similar changes in compliance slope at a particular cycle which is the initiation life of the specimens.

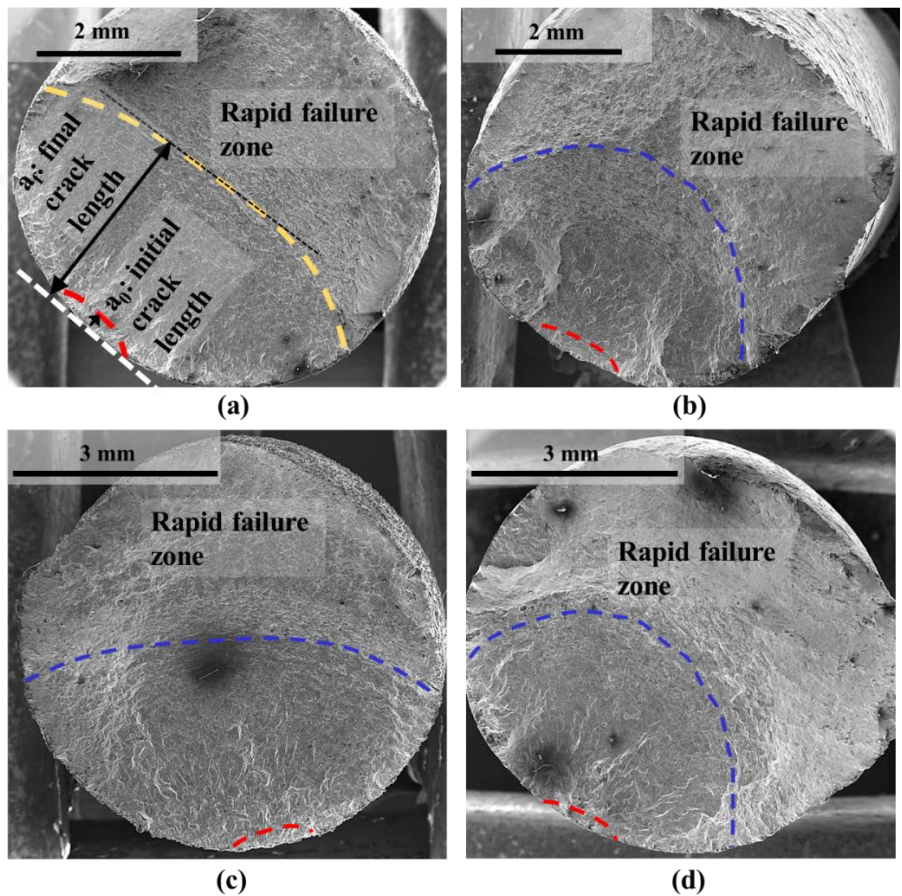

**Figure 8. Fractography images of (a) VR5, (b) VM5, (c) 30R4, and (d) 30M4 specimens depicting the crack initiation zone (red), crack propagation zone (blue), and rapid failure zone.**

The initiation life is also evaluated from fractography images. The fractography image of VR5 in Figure 8(a) reveals the initiation zone, propagation zone, and rapid failure zone. From the initiation zone, the initial crack length is obtained, and the final crack length is calculated perpendicular to the propagation marks till the end of the propagation zone. The initial and final crack lengths are substituted in equation (6) to obtain the specimen's initiation life. The technique is applied to all specimens to evaluate their initiation lives, shown in Figures 8(b), (c), and (d) for representative specimens VM5, 30R4, and 30M4, respectively. A comparison (Figure 9) is presented between the initiation lives evaluated from the compliance technique and fractography analysis. The figure suggests that the results from both methods are in good agreement with each other for all categories of specimens. However, the variation in initiation life is different for different sets of specimens. The VR specimens (Figure 9(a)) show more variation compared to the VM specimens (Figure 9(b)) due to the presence of surface roughness. The 30R specimens (Figure 9(c)) experience the maximum variation in initiation life and the 30M specimens (Figure 9(d)) exhibit similar behavior as the VM specimens by demonstrating small variation. The VR specimens show that a maximum difference of about 10% exists between the two techniques and the difference is for the strain range of 1.5%. As the fatigue life is low at this strain range, a difference of 200 cycles in the initiation lives results in a percentage difference of 10%.

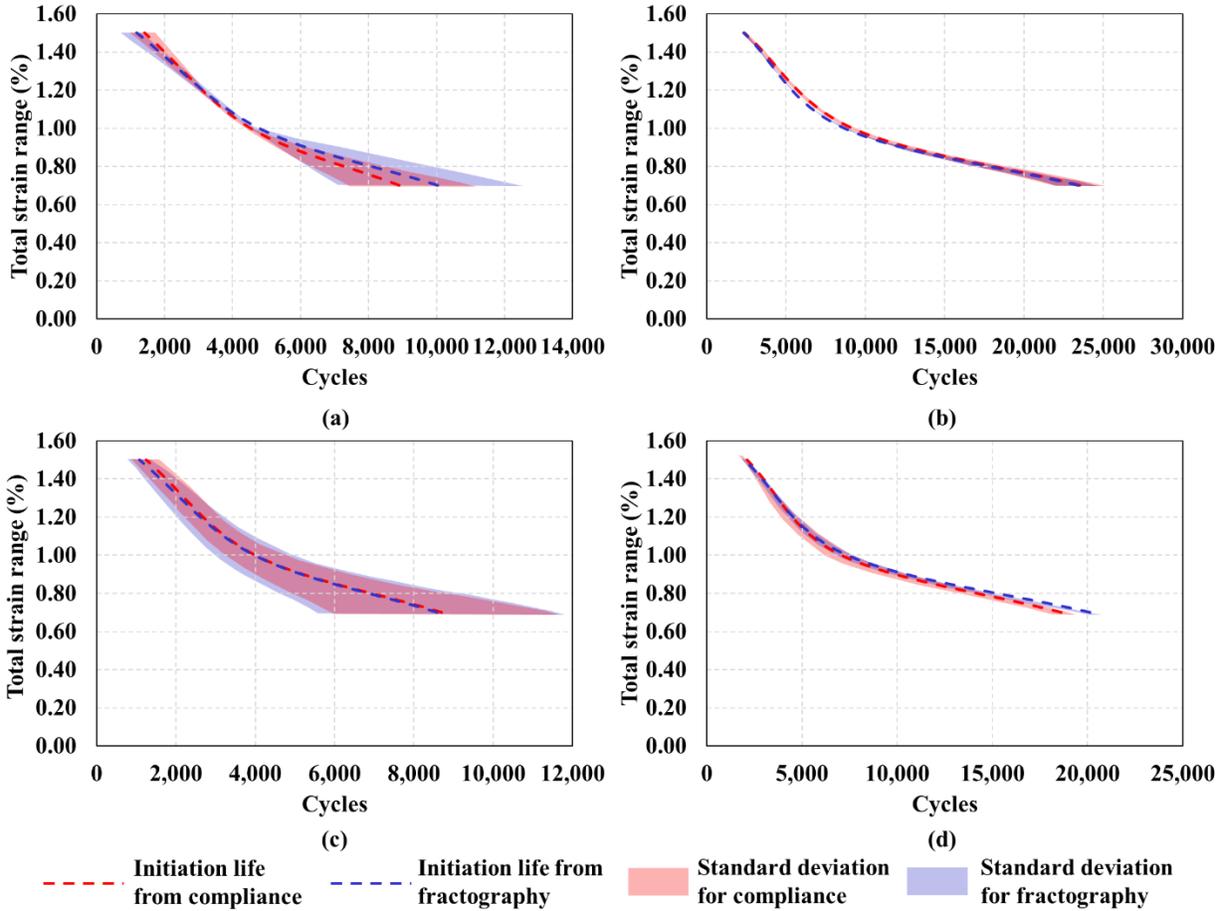

**Figure 9.** Comparison of initiation life results obtained from compliance methodology and fractography analysis: (a) VR, (b) VM, (c) 30R, and (d) 30M. The mean initiation life curves are plotted along with shaded regions depicting the scatter in the results.

## 4 Discussion
### 4.1 Surface roughness analysis

Among all the surface roughness parameters, the maximum valley depth ($R_{v,max}$) is a critical one as it quantifies the valley depth that is expected to act as a notch during fatigue loading. For a particular specimen, surface roughness evaluation can be conducted at multiple longitudinal and transverse cross-sections covering the entire gauge section. The investigation results in multiple $R_{v,max}$ values from each cross-section that is analyzed. The resultant $R_{v,max}$ values must follow a generalized extreme value distribution. This is because the surface roughness profile in L-PBF components follows a normal distribution [68], and a section-wise maximum value of the valley depths yields an extreme value distribution [69]. To measure $R_{v,max}$ values, two pathways are explored as discussed in section 2.1: longitudinal and transverse cross-sectional CT images.

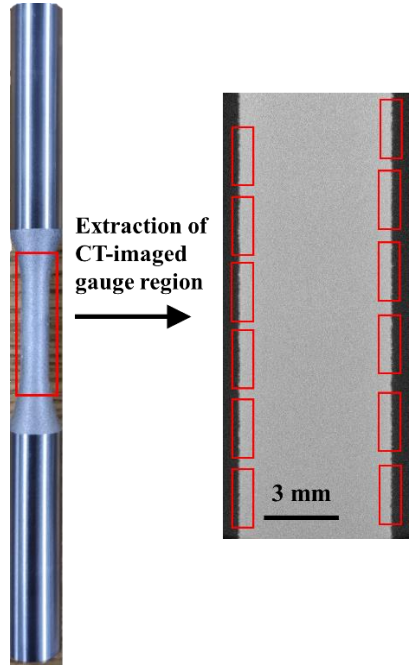

**Figure 10. Selection of multiple zones on the longitudinal cross-sectional CT image of a representative specimen to form the extreme value distribution.**

### 4.1.1 Longitudinal cross-sectional CT image

For longitudinal cross-sectional CT images, 10 to 12 zones are selected that include the specimen's rough surfaces, as shown in Figure 10. These zones are subjected to binarization and edge detection procedures as discussed in section 2.1. A reference rectangle is fitted into the edges to calculate the roughness profiles. From each roughness profile, an $R_{v,max}$ value is obtained, and the $R_{v,max}$ values from all profiles belonging to all zones are used to form the Type 1 Gumbel distribution. To create the distribution, the $R_{v,max}$ values are arranged in ascending order as shown in Table 2, and the parameters G and x are calculated using the following equations [70]:

$$G = \frac{i}{n+1} \qquad (12)$$

$$x = -\ln(-\ln(G)) \qquad (13)$$

Here, $i$ is the corresponding count of a particular $R_{v,max}$ value and $n$ is the total number of zones. The values of $R_{v,max}$, and x are used to plot the distribution as presented in Figure 11. A regression line is fitted to the data with a determination coefficient ($R^2$) value of 0.9694 and the equation of the regression fit is given by,

$$R_{v,max} = 41.021x + 19.024 \qquad (14)$$

From this equation, the distribution mean is calculated using the following relationship,

$$Distribution\ mean = 19.024 + \beta \times 41.021, \quad \beta = 0.577 \qquad (15)$$

$$= 42.69\ \mu m$$

Here, $\beta$ is the Euler-Mascheroni constant and has a value of 0.577. The distribution mean implies the average of $R_{v,max}$ values for a particular specimen (avg($R_{v,max}$)), and the maximum of the distribution value is termed as max($R_{v,max}$). This process is carried out for all the *rough* specimens to evaluate avg($R_{v,max}$) and max($R_{v,max}$) values.

**Table 2. Distribution of $R_{v,max}$ with parameters of the Gumbel distribution for longitudinal cross-sectional CT images for one specimen. Each serial number (S. no.) corresponds to a CT cross-sectional image**

| S. no. (i) | $R_{v,max}$ ($\mu m$) | G=i/(n+1) | x=-ln(-ln(G)) |
|---|---|---|---|
| 1 | 22 | 0.1 | 0 |
| 2 | 23.46667 | 0.2 | 0.155541 |
| 3 | 29.33333 | 0.3 | 0.281599 |
| 4 | 33 | 0.4 | 0.400182 |
| 5 | 44 | 0.5 | 0.52139 |
| 6 | 45.46667 | 0.6 | 0.653943 |
| 7 | 54.26667 | 0.7 | 0.809943 |
| 8 | 55 | 0.8 | 1.013631 |
| (n) 9 | 77 | 0.9 | 1.339538 |

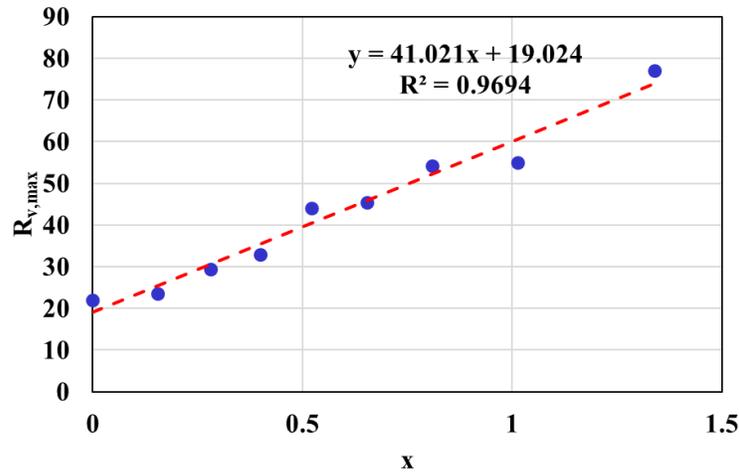

**Figure 11. The regression line (red) fits onto the $R_{v,max}$ values (blue), which are obtained from the longitudinal cross-section image. The regression line equation along with the $R^2$ value is shown in the image.**

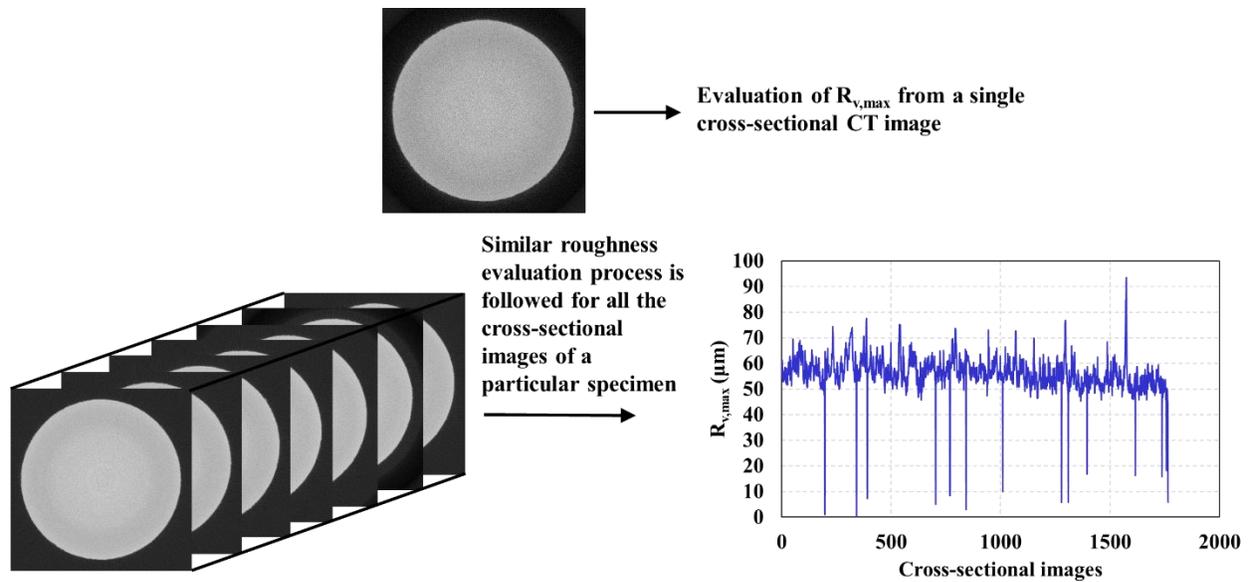

**Figure 12.** Extraction of $R_{v,max}$ value from a single transverse cross-sectional CT image along with the extraction from all the images of a particular specimen. The variation of $R_{v,max}$ with images depicts the scatter in the roughness data in one representative specimen which justifies the necessity to create a distribution with the data.

**Table 3.** Distribution of $R_{v,max}$ with parameters of Gumbel distribution for transverse cross-sectional CT images for one representative specimen. Each serial number (S. no.) corresponds to a CT cross-sectional image.

| S. no. (i) | $R_{v,max}$ (um) | G=i/(n+1) | x=-ln(-ln(G)) |
|---|---|---|---|
| 1 | 29.70124 | 0.000571 | -2.01054 |
| 2 | 29.70124 | 0.001143 | -1.91312 |
| 3 | 29.79533 | 0.001714 | -1.8514 |
| 4 | 29.92707 | 0.002286 | -1.80518 |
| 5 | 29.98353 | 0.002857 | -1.7678 |
| 6 | 30.12469 | 0.003429 | -1.73618 |
| 7 | 30.15292 | 0.004 | -1.70864 |
| 8 | 30.3035 | 0.004571 | -1.68416 |
| 9 | 30.3882 | 0.005143 | -1.66206 |
| 10 | 30.4729 | 0.005714 | -1.64186 |

### 4.1.2 Transverse cross-sectional CT image

Transverse cross-sectional CT images are used to compare the valley depths with the longitudinal cross-sectional images. If the roughness valley dimensions are comparable in both directions, the valleys can be considered as semi-ellipsoidal notches as discussed in the next paragraph. For one specimen, 1,600 to 1,800 transverse cross-sectional CT images are used to evaluate the roughness profile. Each cross-sectional CT image undergoes binarization and edge detection as described in section 2.1. A circle of diameter 6 mm is fitted on the rough circular edge and the valley depth is calculated as shown in Figure 4. All the transverse cross-sectional images of a specimen are used to obtain the $R_{v,max}$ values as shown in Figure 12. The $R_{v,max}$ values are arranged in ascending order and the first ten values are presented in Table 3. Using these values, the parameters G and x are evaluated from equations (12) and (13), respectively, and they are tabulated in Table 3. The $R_{v,max}$ values are plotted against x in Figure 13, and a regression line is fitted on the 1,700 data points. From the regression fit, an equation is established between $R_{v,max}$, and x:

$$R_{v,max} = 3.1091x + 35.936 \quad (16)$$

The distribution mean is calculated from the above equation using the relationship in equation (15) and avg($R_{v,max}$) is 37.72 $\mu m$. Similarly, avg($R_{v,max}$) and max($R_{v,max}$) are calculated for all the *rough* specimens.

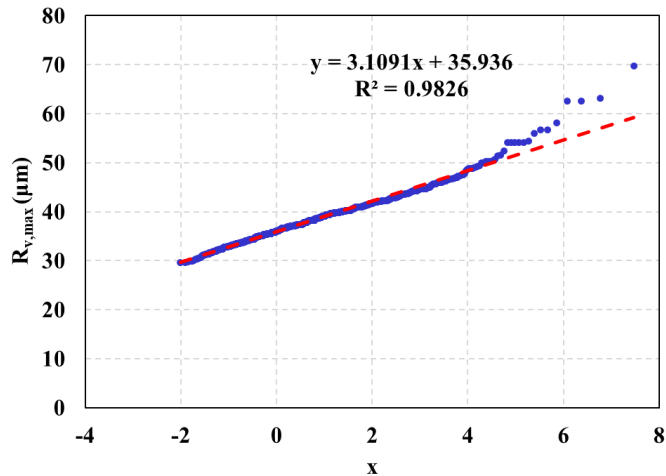

**Figure 13. The regression line (red) fits onto the $R_{v,max}$ values (blue), which are obtained from the transverse cross-section images. The regression line equation along with the $R^2$ value is shown in the image.**

A comparison between avg($R_{v,max}$) values obtained from longitudinal and transverse cross-sectional CT images of different *rough* specimens is shown in Figure 14. The comparison suggests that both cross-sectional images result in similar maximum notch depths, and the results from longitudinal cross-sectional images can be used for developing functional relationships. Longitudinal results of surface roughness are more important than transverse cross-sectional images because the fatigue loading is in the longitudinal direction, and the opening of the notches is going to be in a direction perpendicular to the loading direction. Moreover, the comparison suggests that the notches are semi-ellipsoidal as the longitudinal and transverse cross-sectional images provide similar roughness values. This implies that equations and correlations of semi-elliptical notches can be applied to these *rough* specimens.

From the longitudinal cross-sectional CT images, the roughness profile along a longitudinal length of the rough edges is obtained as shown in Figures 15(a) and (b). The roughness profile showcases global peaks and valleys along the specimen's length. The maximum notch depth in each longitudinal image resides within the global valley of the surface. The distance between two such valleys along the circumference is termed mean waviness spacing ($R_{sm}$) as shown in Figure 15(b). The valley regions are prone to crack initiation during fatigue loading as those regions give rise to stress concentration. Hence, the difference in fatigue life of the *rough* and *machined* specimens lies in the surface roughness of the *rough* specimens. Therefore, a functional relationship is required to understand the impact of surface roughness on the low-cycle fatigue life of a component. Furthermore, the analogous notch dimensions are depicted in Figure 15(c) which are utilized in developing the functional relationship.

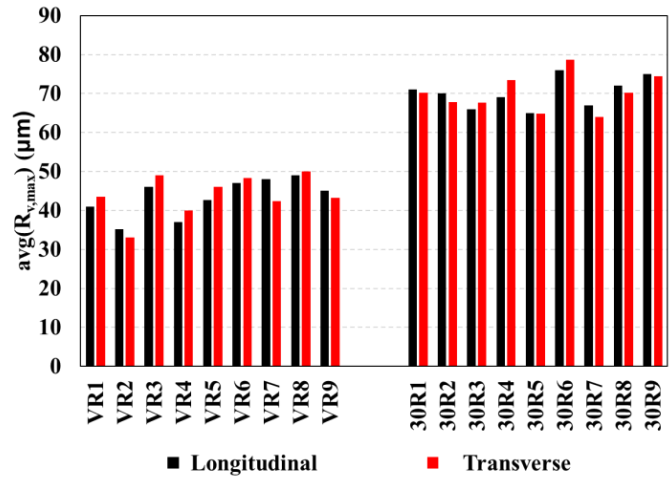

Figure 14. Comparison between avg($R_{v,max}$) values obtained from longitudinal and transverse cross-sectional CT images, and the comparison shows that both methods result in similar values.

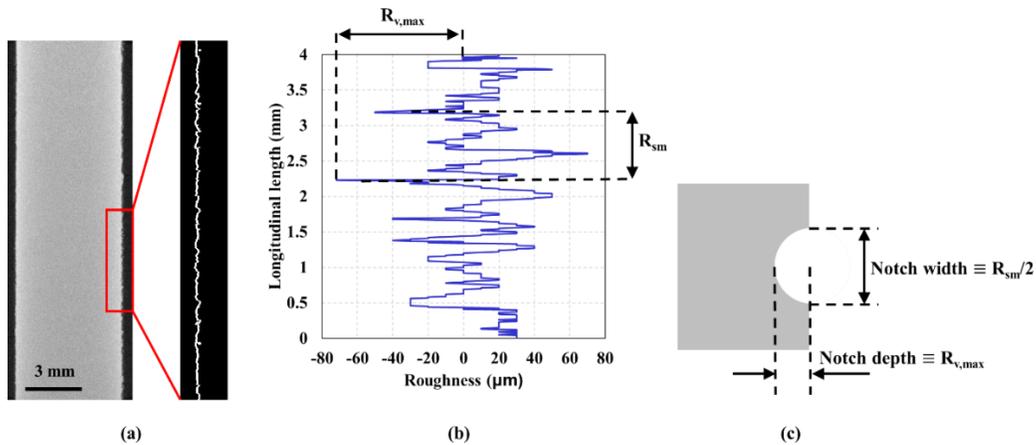

Figure 15. (a) Extraction of roughness peaks and valleys from the CT image shows the (b) variation of roughness along the edge length. (c) A schematic representation of a notch showing the notch depth and notch width equivalence to $R_{v,max}$, and $R_{sm}$, respectively.

### 4.2 Fatigue behavior evaluation
#### 4.2.1 Load-cycle variation during fatigue

From the fatigue life investigation, load-cycle and stress-strain hysteresis plots are obtained for all specimens. For each specimen set, three load-cycle curves are shown at three different strain ranges, presented in Figure 16. In the VR specimen set (Figure 16(a)), load-cycle curves correspond to VR3, VR6, and VR8, tested at 1.5%, 1%, and 0.7% total strain ranges, respectively. The variation of load is significantly different for each strain range. At the 1.5% strain range, VR3 shows a sudden increase in load during the first few cycles, which implies steep strain hardening during those cycles. After reaching the peak load close to 12,000 N, the load continues to be around 11,500 N throughout the entire fatigue life. On the other hand, the load variation is different for VR6 which is investigated at a total strain range of 1%. The load gradually increases with the number of cycles, unlike the steep rise shown in VR3. It implies that the strain hardening behavior is gradual at 0.5% strain amplitude. Moreover, the peak load is 10,500 N which is less than the peak load of 12,000 N at a 1.5% strain range. The peak load is even less in the case of the 0.7% total strain range for VR8. At this strain range, the load increases very slowly with the increase in cycles till the specimen reaches around 8,000 cycles, and then the load gradually decreases with the increase in cycles until the failure of the specimen. However, the slope of load-cycle curves is similar at 1% and 0.7% total strain ranges. The load variation shows a similar trend, but the peak load values are different for the two cases.

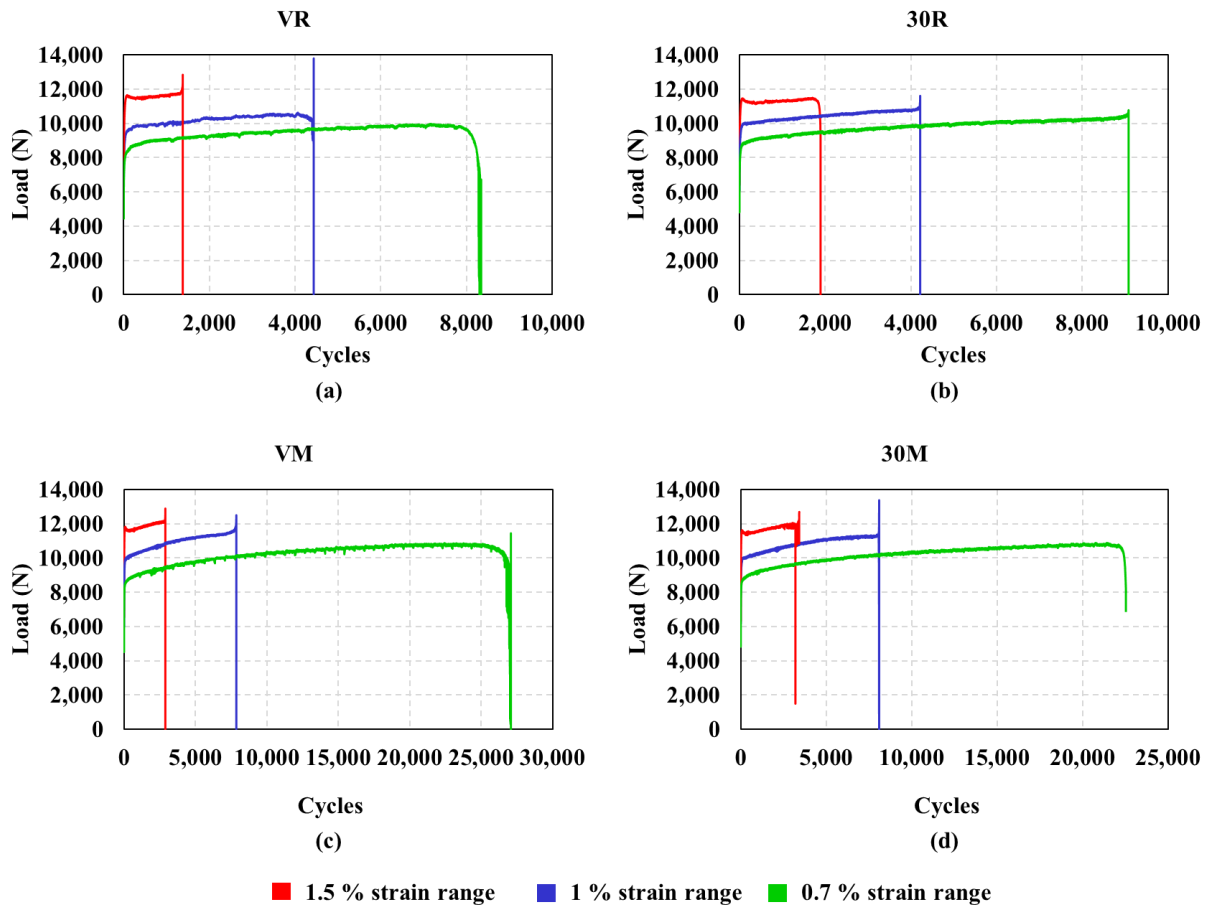

**Figure 16. Load-cycle curves of (a) VR, (b) 30R, (c) VM, and (d) 30M. Each set presents load-cycle variation at 3 strain ranges.**

Similar load variation plots are created for the other sets of specimens tested at different strain ranges. The 30R specimens (Figure 16(b)) show a load variation similar to the VR specimens. The nature of strain hardening, and the peak load values are close to the VR specimens for the respective strain ranges. On the contrary, *machined* (Figures 16(c) and (d)) specimens show a slightly different behavior than the *rough* specimens. The *machined* specimens encounter a higher amount of strain hardening compared to the *rough* specimens. At the 1.5% strain range, VM3 and 30M3 present a steep strain hardening in the first few cycles followed by a decrease in load for a very few cycles, and then the load increases till the specimen fails. Therefore, the strain hardening is much greater in the *machined* specimens with a peak load of 12,000 N. Also, at the other strain ranges of 1% and 0.7%, the *machined* specimens undergo elevated strain hardening because they can endure a higher load than the *rough* specimens. For example, at the strain range of 0.7%, it can be observed that VR8 and 30R8 reach the peak load of around 10,000 N at 7,153 cycles. However, VM8 and 30M8 reach similar load values at 7,153 cycles, which sheds insight into the fact that material behavior is similar across all sets of specimens, but the *rough* specimens fail early. Hence, it can be inferred from the load variation plots of *rough* and *machined* specimens that the former experiences fracture or final failure significantly earlier than the cycle at which the latter reaches their peak load. It suggests the surface roughness has a significant contribution in determining the specimens' low cycle fatigue life.

### 4.2.2 Stress-strain hysteresis behavior

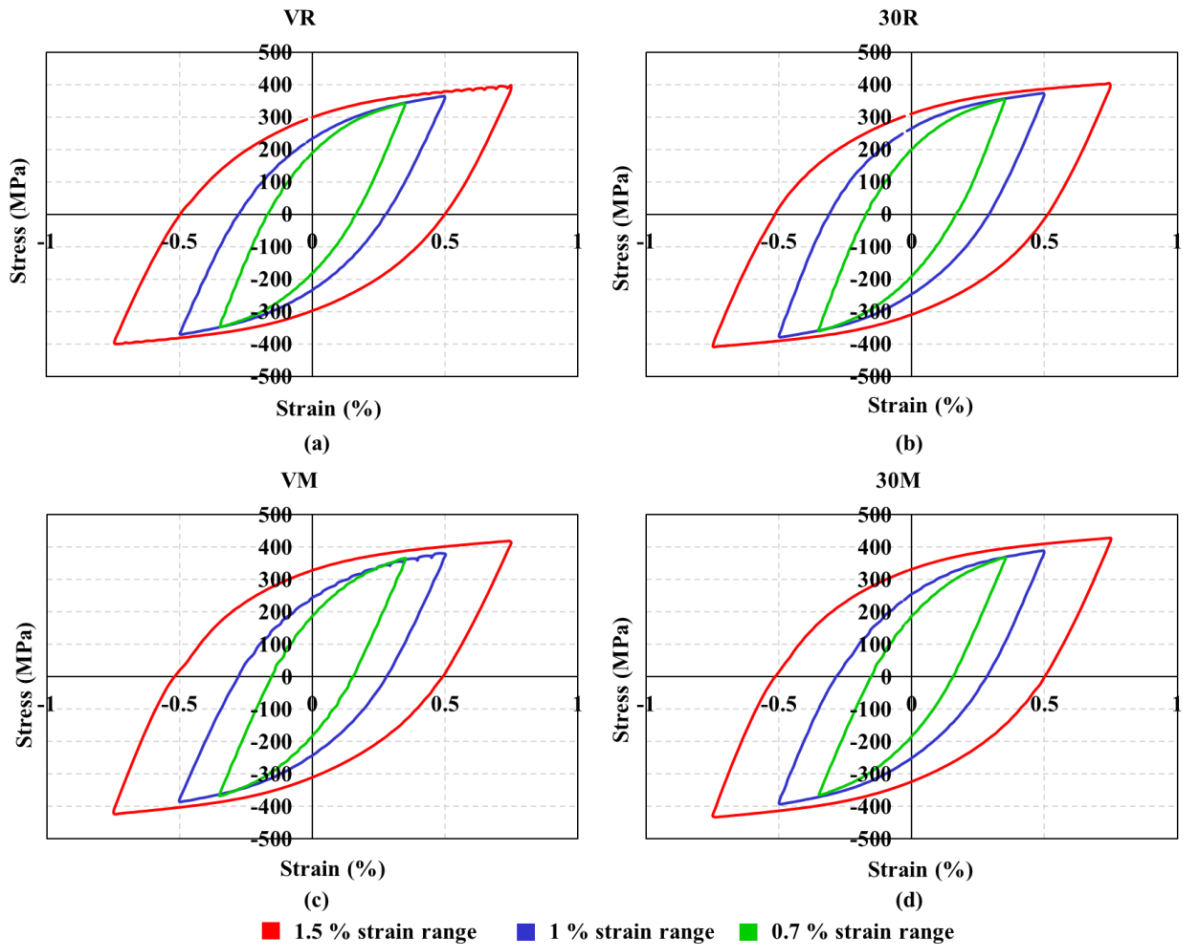

Figure 17. Stable cycle stress-strain hysteresis curves of (a) VR, (b) 30R, (c) VM, and 30M. Each set presents stress-strain hysteresis curves at 3 strain ranges.

Apart from the load-cycle variation, stress-strain hysteresis curves at different strain ranges provide the cyclic stress-strain curve, shown in Figure 17. The stable cycle or the half-life hysteresis curves are plotted for the VR specimens in Figure 17(a). The stress-strain curves reinforce the observations gathered from the load-cycle variation. The specimen tested at the 1.5% strain range showcases a stress-strain hysteresis loop with a larger enclosed area followed by the hysteresis loops of the specimens tested at 1% and 0.7% strain ranges, respectively. The maximum stress in the tensile region decreases with a decrease in strain range. A similar trend is observed for the 30R specimens (Figure 17(b)), and the maximum stresses at each strain range are close to the stresses observed for VR specimens. However, *machined* specimens (Figures 17(c) and 17(d)) show differences in stress values compared to *rough* specimens. At the 1.5% strain range, all VM and 30M specimens endure a stress value of 420 MPa at half-life whereas the VR and 30R specimens encounter a stress of 390 MPa at half-life. As the *rough* specimens undergo early fatigue failure, their half-lives are less than the *machined* specimens as observed from the *load-cycle* curve. The early fatigue failure is also corroborated by fractography analysis.

### 4.3 Initiation life evaluation
#### 4.3.1 Compliance method

The compliance methodology is based on the load-displacement data obtained from the fatigue investigation of the specimens. The application of an actuator displacement to a specimen is analogous to a force on a spring, where the spring stiffness is the specimen stiffness. After several loading cycles, when a crack initiates in a specimen, the specimen's stiffness decreases due to the reduction in cross-sectional gauge area by following the equation,

$$K = \frac{EA}{l} \tag{7}$$

Here, K is the specimen's stiffness, E is the elastic modulus, A is the gauge cross-section area, and $l$ is the reference gauge length. With the reduction in gauge area, the specimen's stiffness decreases which in turn increases the specimen's compliance as it is the inverse of stiffness, given by the relation,

$$C = \frac{1}{K} \tag{8}$$

As stiffness can also be described from spring analogy by the following relation,

$$K = \frac{F}{D} \tag{9}$$

Here, compliance can be described as $C = \frac{1}{K} = \frac{D}{F}$. During the fatigue investigation, strain hardening causes an increase in actuator displacement to maintain the same strain range. The increase in actuator displacement results in a steady increase in compliance as shown in Figure 18(a). From the compliance-displacement variation curve, it is challenging to identify the crack initiation in the specimen. However, the evaluation of the compliance-displacement slope from the curve pinpoints the crack initiation instance in the specimen by showing a marked increase in the slope at the crack initiation cycle. This is because a crack initiation is always associated with a load drop and the load drop is more pronounced with the evaluation of the compliance-displacement slope. Suppose for an $i^{th}$ cycle, compliance is $C_i$, displacement is $D_i$, and load is $F_i$. If the crack initiates at $i+1^{th}$ cycle, load $F_{i+1}$ is less than $F_i$ but $D_{i+1}$ remains close to the value $D_i$ as the actuator displacement maintains the same strain range. Therefore, $C_{i+1}$ is greater than $C_i$ increasing the compliance-displacement slope, given by

$$Slope = \frac{C_{i+1} - C_i}{D_{i+1} - D_i}. \tag{10}$$

As the difference between $D_{i+1}$ and $D_i$ is much less compared to the difference between $C_{i+1}$ and $C_i$, the slope denotes a marked increase at the crack initiation cycle as presented in Figure 18(b). This methodology is applied to all the fatigue investigations to evaluate the initiation life of the specimens and the result is further corroborated by initiation life evaluation from fractography analysis as already shown in section 3.3. The initiation life values obtained from compliance methodology are employed to form an empirical relationship with the fatigue life of the specimens as depicted in Figure 18(c). On the experimental observations in the figure, a power curve is fitted with an $R^2$ value of 0.97. The power curve describes the relationship between initiation life and fatigue life as,

$$N_i = 0.75 N_f^{1.015} \tag{11}$$

Here, $N_i$ is the initiation life and $N_f$ is the fatigue life of a specimen. The relationship suggests that the initiation life of an L-PBF HX specimen is approximately 75% of the fatigue life of the specimen, which corroborates with the observation in the open literature [71].

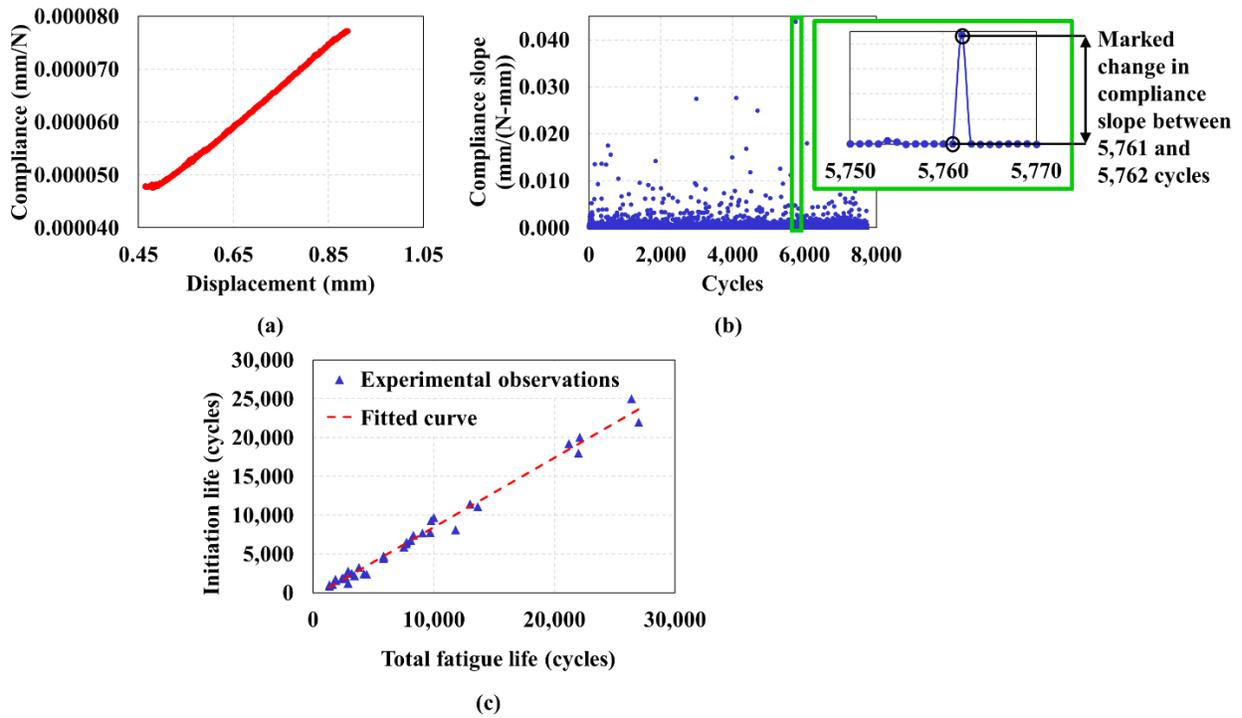

**Figure 18. (a) Variation of compliance with displacement for specimen VM6 tested at 1% total strain range, (b) variation of the compliance-displacement slope with loading cycles, depicting a marked increase in the slope at 5,762 cycles, and (c) variation of initiation life with total fatigue life of all specimens.**

### 4.3.2 Fractography analysis

Following the trend of fatigue failure of all the specimens, the initiation life of the specimen also shows a similar trend. In *rough* specimens, surface roughness plays a critical role in dictating the failure of the specimens because surface valleys act as notches during fatigue loading. At a particular cycle, a crack originates from a surface feature which may become the primary crack leading to the failure. The convergence of ratchet marks on the fractured surface indicates the crack initiation region, shown in the fractography analysis of VR5 (Figure 19(a)). In this figure, different regions can be identified as crack initiation sites from the pattern of ratchet marks [72] as observed on the fractured surface (Figure 19(a)). The primary crack can be distinguished from the secondary cracks by observing the crack propagation or beach marks (Figure 19(b)). The primary crack growth direction is perpendicular to the crack propagation marks which implies that the crack initiation location, marked by a red dashed line in Figure 19(c), is the origination of the primary crack [73]. On the other hand, VR3 (1.5% strain range) and VR8 (0.7% strain range) specimens in Figures 20(a) and 20(b), respectively, show multiple cracks on the specimens' surfaces. However, the cracks start to grow in a favored slip plane direction where there are no hindrances from dislocations [74]. Therefore, the primary crack growth is favored by a single region where the dislocations are the least in amount, or in other words, where the dislocations have favored the crack extension leading to failure [75].

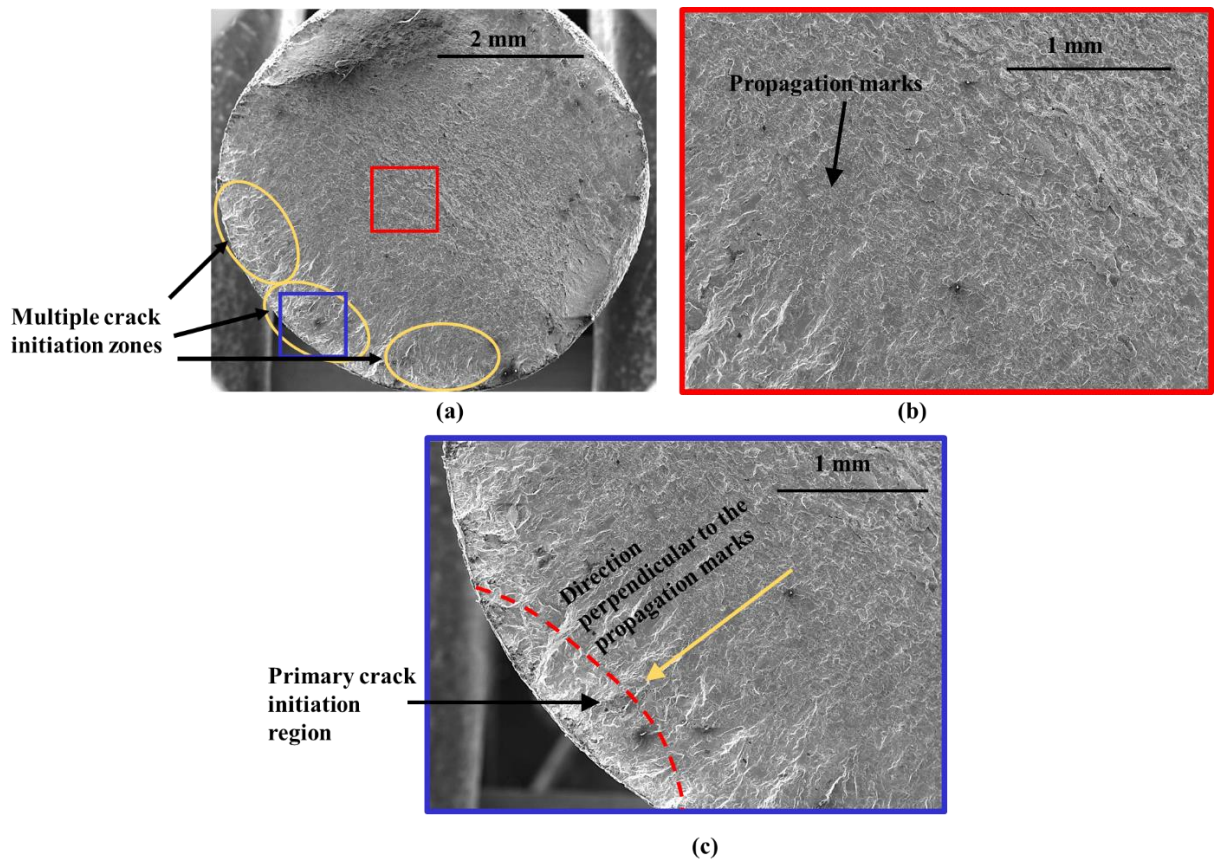

**Figure 19. Fractography images of the VR5 specimen tested at a 1% strain range. The zoomed-in images of (a) and (b) reveal the propagation marks on the fractured surface and the primary crack initiation region is identified by moving in a perpendicular direction to the propagation marks. The whole fractography image in (c) shows the presence of multiple crack initiation zones on the fractured surface.**

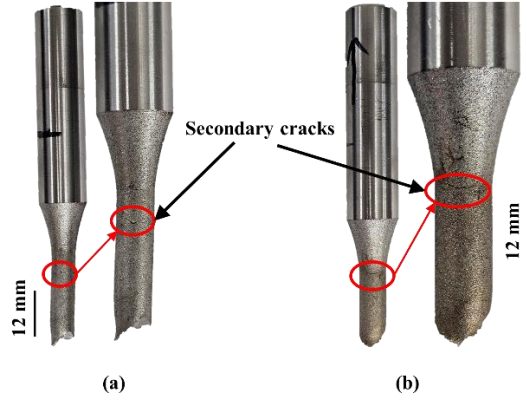

**Figure 20. Presence of secondary cracks on the rough surfaces of (a) VR3 and (b) VR8 specimens. Both these specimens belong to the *rough* category.**

### 4.4 Functional relationship between surface roughness and fatigue life

As discussed in section 4.1, the surface valleys are semi-elliptical notches where the maximum valley depth of each specimen is used as the notch depth, and the half of R$_{sm}$ between two surface valleys is considered as the notch width. For a specimen, there is a maximum notch depth assumed to be the crack initiation location. However, the crack originating from maximum notch depth may not be the primary crack leading to failure. Therefore, a distribution of maximum notch depths is obtained from longitudinal cross-sectional CT images. The maximum value distribution corresponds to the Gumbel distribution from which the avg(R$_{v,max}$) is obtained for a particular specimen, as described in section 4.1.1. The following assumptions are considered while deriving the relationship between surface roughness and fatigue life: (i) surface notches are semi-elliptical notches and (ii) uniaxial stress is applied in a perpendicular direction to the notch plane (importance of longitudinal cross-sectional images). The assumption of semi-elliptical notches is justified by the similar avg(R$_{v,max}$) values obtained from longitudinal and transverse cross-sectional CT images. These assumptions bring out the relationship between local stress around a notch, which is given by,

$$\sigma_{max} = \sigma_{nom}\left(1 + \frac{2a}{b}\right) \tag{17}$$

Here, a is the notch depth, b is the notch radius and $\sigma_{nom}$ is the nominal stress that the specimen is subjected to during fatigue loading. In the present scenario, a is avg(R$_{v,max}$) and b is R$_{sm}$/2, which leads to the equation:

$$\sigma_{max} = \sigma_{nom}\left(1 + \frac{4 \times avg(R_{v,max})}{R_{sm}}\right) \tag{18}$$

The maximum local stress, obtained from the equation, is applied to the Ramberg-Osgood equation to determine the maximum local plastic strain, as described in the following equation:

$$\epsilon_{p,notch} = \left(\frac{\sigma_{max}}{K'(T)}\right)^{\frac{1}{n'(T)}} \tag{19}$$

Here, $K'(T)$ is the cyclic strength coefficient and $n'(T)$ is the cyclic strain hardening exponent. The maximum plastic strain value implies the strain localization around a surface notch, and it is employed in the Coffin-Manson-Basquin relation to predict the fatigue life of the *rough* specimens. The relationship between plastic strain and fatigue life is given by,

$$\frac{\Delta \epsilon_p}{2} = \epsilon_f'(T) \times (N_f)^{c(T)} \tag{20}$$

Here, $\Delta \epsilon_p/2$ is the plastic strain amplitude, $\epsilon_f'(T)$ is the fatigue ductility coefficient, $c(T)$ is the fatigue ductility exponent, and $N_f$ is the total fatigue life. The temperature-dependent material parameters such as cyclic strain hardening exponent, fatigue ductility coefficient, and fatigue ductility exponent are adopted from the literature [76], [77]. The plastic strain amplitude is substituted by $\epsilon_{p,notch}$ from the Ramberg-Osgood relation, which results in the equation:

$$\left(\frac{\sigma_{max}}{K'(T)}\right)^{\frac{1}{n'(T)}} = \epsilon_f'(T) \times (N_f)^{c(T)} \tag{21}$$

Here, $\sigma_{max}$ is substituted by equation (17) which results in the following relationship:

$$\left(\frac{\sigma_{nom}}{K'(T)}\left(1 + \frac{4 \times avg(R_{v,max})}{R_{sm}}\right)\right)^{\frac{1}{n'(T)}} = \epsilon_f'(T) \times (N_f)^{c(T)} \tag{22}$$

Here, $\left(\frac{\sigma_{nom}}{K'(T)}\right)^{\frac{1}{n'(T)}}$ is substituted as $\epsilon_{p,nom}$, which is the applied plastic strain amplitude in an experiment. The final equation is given by,

$$\epsilon_{p,nom}\left(1 + 4 \times \frac{avg(R_{v,max})}{R_{sm}}\right)^{\frac{1}{n'(T)}} = \epsilon_f'(T) \times (N_f)^{c(T)} \tag{23}$$

The known parameters in the equation are the plastic strain amplitude to which an L-PBF specimen is subjected during the fatigue investigation, the temperature-dependent material parameters, and the surface roughness parameters, avg($R_{v,max}$) and $R_{sm}$. The roughness parameters avg($R_{v,max}$) and $R_{sm}$ are evaluated for all specimens from the transverse cross-sectional CT images. The extraction of material properties and the evaluation of the roughness parameters are followed by the prediction of fatigue lives of *rough* specimens. The fatigue lives of all *rough* specimens are predicted using equation (23) and a mean curve is obtained for each specimen set. A lower bound fatigue curve is obtained by substituting max($R_{v,max}$) as avg($R_{v,max}$) in equation (23) for all *rough* specimens and the mean curve is extracted for each specimen set.

The predicted fatigue life values of different specimens are compared with the experimental fatigue life values, shown in Figure 21. For the VR specimens, a mean fatigue life curve is predicted from the equation and presented in Figure 21(a) with a shaded region showing the standard deviation of the prediction. The mean fatigue life curve from the equation overlaps the mean fatigue life values from the experiments, resulting in a maximum difference of 2%. The lower bound fatigue curve in Figure 21(b) shows a difference from the experimental curve at lower strain ranges, which implies that if the primary crack grows from the worst surface notch, the fatigue life will be less compared to the experimental observations. A crack may initiate from the worst surface notch, but the crack growth is determined by the grain boundaries and dislocation pile-ups. Due to these uncertainties, the variance of the mean fatigue life curve of avg($R_{v,max}$) is shown by the shaded region.

In Figure 21(c), 30R specimens show a difference between the experimental observations and the mean fatigue life curve from avg($R_{v,max}$). The maximum difference between the mean fatigue life curve from the equation and experimental mean values is around 10%, occurring at the lowest strain range. This difference at low strain range is due to the influence of build direction along with surface roughness [38]. The mean fatigue life curve from the worst surface notch or the max($R_{v,max}$) shows fatigue life values that are significantly less than the experimental observations, presented in Figure 21(d). This implies that the maximum notch depth on the specimens' down-skin surfaces may lead to a primary crack and premature fatigue failure. However, the crack growth is dependent on the accumulation of dislocations at the grain boundary due to which fatigue cracks may originate from other valleys and lead to the fatigue failure of a specimen. To incorporate the probability of the worst surface notch not leading to the specimen's failure, the mean life curve from avg($R_{v,max}$) is found to be more prolific in determining the operational life.

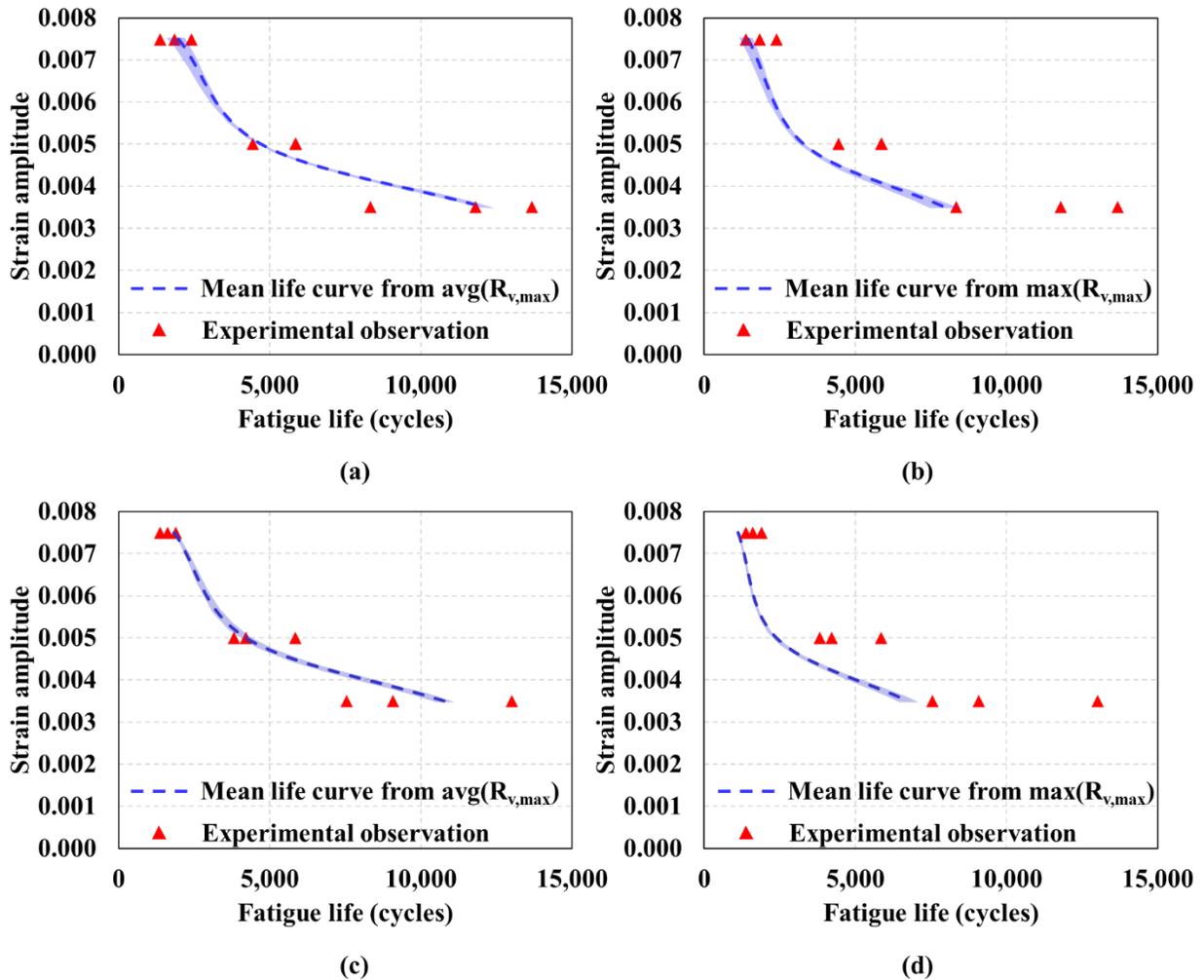

Figure 21. (a) Mean life curve of VR specimens predicted from equation (23) using avg($R_{v,max}$), (b) Mean life curve of VR specimens predicted from equation (23) using max($R_{v,max}$), (c) Mean life curve of 30R specimens predicted from equation (23) using avg($R_{v,max}$), and (d) Mean life curve of 30R specimens predicted from equation (23) using max($R_{v,max}$). max($R_{v,max}$) values indicate the lower limit of the fatigue curve of a specimen. No specimens exhibit fatigue life less than the precited lower bound fatigue curve as shown in (b) and (d). Blue-shaded regions indicate the standard deviation of the predicted curves.

5. Conclusion

In this work, fatigue investigations of L-PBF HX specimens are carried out at an elevated temperature of 500 °F. Four different sets of specimens: VR, VM, 30R, and 30M, are used to understand the impact of surface roughness on the fatigue properties. Cross-sectional CT images are used to evaluate and create a

Gumbel distribution of maximum valley depths of the *rough* specimens. The average value and the maximum value of the distribution of each specimen are used to build a functional relationship between the strain range, material parameters, valley depth, and fatigue life of the specimens. The following key conclusions can be inferred from the work:

- 30R specimens consist of up-skin and down-skin surfaces, and the down-skin surfaces have higher roughness. On the other hand, VR specimens have consistent surface roughness throughout the gauge section.
- During fatigue investigations, the *rough* specimens fail early compared to the *machined* specimens because the surface valleys in the *rough* specimens act as notches that lead to strain localization and fatigue failure.
- Crack initiation is always associated with load-drop, however, the initiation phenomena can be more accurately characterized through compliance. The variation of the compliance-displacement slope with loading cycles highlights the probable initiation life.
- Fractography reveals crack initiation and crack propagation zones, which are used to evaluate the initiation life of the specimens, using Paris law coefficients. The initiation life evaluation from compliance methodology and fractography analysis suggest that both techniques are reliable for pinpointing the initiation life of L-PBF specimens.
- The functional relationship predicts a mean fatigue life and a minimum operational life of L-PBF specimens with surface roughness. The mean life curve from the functional relationship shows a maximum difference of 2% from the experimental mean fatigue life observations for VR specimens and 10% for 30R specimens.

The functional relationship can be applied to other metal AM processes such as laser-directed energy-deposition and binder jetting. Furthermore, the analytical methodology can be extended to investigations involving in-phase and out-of-phase thermomechanical fatigue. In this work, the predictions correspond to the low cycle fatigue life of L-PBF specimens, and in the future, fatigue investigations can be conducted to observe the prediction accuracy of the functional relationship in the high cycle fatigue regime. Moreover, the analytical methodology can be combined with numerical and data-driven models to improve the accuracy of predictions.

**CRediT authorship contribution statement**

**Ritam Pal**: Investigation, Formal Analysis, Methodology, Validation, Software, Visualization, Writing - Original Draft. **Brandon Kemerling**: Supervision, Writing - Review and Editing. **Daniel Ryan**: Supervision, Writing - Review and Editing. **Sudhakar Bollapragada**: Supervision, Writing - Review and Editing. **Amrita Basak**: Conceptualization, Methodology, Resources, Supervision, Funding Acquisition, Writing - Review and Editing.

**Funding**

The work reported in this paper is funded by Solar Turbines Inc., a Caterpillar Company through grant number PAT-18-00112.

**Acknowledgments**

The authors would like to thank Metcut Research Inc. (Cincinnati, Ohio) and the Center for Quantitative Imaging (Penn State) for their help with machining and computed tomography, respectively.

# References

[1] L. Hirt, A. Reiser, R. Spolenak, and T. Zambelli, "Additive Manufacturing of Metal Structures at the Micrometer Scale," *Advanced Materials*, vol. 29, no. 17. Wiley-VCH Verlag, May 03, 2017. doi: 10.1002/adma.201604211.

[2] J. O. Milewski, "Springer Series in Materials Science 258 Additive Manufacturing of Metals From Fundamental Technology to Rocket Nozzles, Medical Implants, and Custom Jewelry." [Online]. Available: http://www.springer.com/series/856

[3] D. Svetlizky *et al.*, "Directed energy deposition (DED) additive manufacturing: Physical characteristics, defects, challenges and applications," *Materials Today*, vol. 49. Elsevier B.V., pp. 271–295, Oct. 01, 2021. doi: 10.1016/j.mattod.2021.03.020.

[4] S. L. Sing and W. Y. Yeong, "Laser powder bed fusion for metal additive manufacturing: perspectives on recent developments," *Virtual and Physical Prototyping*, vol. 15, no. 3. Taylor and Francis Ltd., pp. 359–370, Jul. 02, 2020. doi: 10.1080/17452759.2020.1779999.

[5] M. Ziaee and N. B. Crane, "Binder jetting: A review of process, materials, and methods," *Additive Manufacturing*, vol. 28. Elsevier B.V., pp. 781–801, Aug. 01, 2019. doi: 10.1016/j.addma.2019.05.031.

[6] S. Chowdhury *et al.*, "Laser powder bed fusion: a state-of-the-art review of the technology, materials, properties & defects, and numerical modelling," *Journal of Materials Research and Technology*, vol. 20. Elsevier Editora Ltda, pp. 2109–2172, Sep. 01, 2022. doi: 10.1016/j.jmrt.2022.07.121.

[7] J. Elambasseril, J. Rogers, C. Wallbrink, D. Munk, M. Leary, and M. Qian, "Laser powder bed fusion additive manufacturing (LPBF-AM): the influence of design features and LPBF variables on surface topography and effect on fatigue properties," *Critical Reviews in Solid State and Materials Sciences*, vol. 48, no. 1. Taylor and Francis Ltd., pp. 132–168, 2023. doi: 10.1080/10408436.2022.2041396.

[8] A. Hadadzadeh, B. S. Amirkhiz, S. Shakerin, J. Kelly, J. Li, and M. Mohammadi, "Microstructural investigation and mechanical behavior of a two-material component fabricated through selective laser melting of AlSi10Mg on an Al-Cu-Ni-Fe-Mg cast alloy substrate," *Addit Manuf*, vol. 31, Jan. 2020, doi: 10.1016/j.addma.2019.100937.

[9] A. Keshavarzkermani, M. Sadowski, and L. Ladani, "Direct metal laser melting of Inconel 718: Process impact on grain formation and orientation," *J Alloys Compd*, vol. 736, pp. 297–305, Mar. 2018, doi: 10.1016/j.jallcom.2017.11.130.

[10] A. Keshavarzkermani *et al.*, "An investigation into the effect of process parameters on melt pool geometry, cell spacing, and grain refinement during laser powder bed fusion," *Opt Laser Technol*, vol. 116, pp. 83–91, Aug. 2019, doi: 10.1016/j.optlastec.2019.03.012.

[11] U. Ali *et al.*, "On the measurement of relative powder-bed compaction density in powder-bed additive manufacturing processes," *Mater Des*, vol. 155, pp. 495–501, Oct. 2018, doi: 10.1016/j.matdes.2018.06.030.

[12]  U. Ali *et al.*, "Identification and characterization of spatter particles and their effect on surface roughness, density and mechanical response of 17-4 PH stainless steel laser powder-bed fusion parts," *Materials Science and Engineering: A*, vol. 756, pp. 98–107, May 2019, doi: 10.1016/j.msea.2019.04.026.

[13]  A. Keshavarzkermani *et al.*, "Controlling mechanical properties of additively manufactured hastelloy X by altering solidification pattern during laser powder-bed fusion," *Materials Science and Engineering: A*, vol. 762, Aug. 2019, doi: 10.1016/j.msea.2019.138081.

[14]  Y. Mahmoodkhani *et al.*, "On the measurement of effective powder layer thickness in laser powder-bed fusion additive manufacturing of metals," *Progress in Additive Manufacturing*, vol. 4, no. 2, pp. 109–116, Jun. 2019, doi: 10.1007/s40964-018-0064-0.

[15]  J. D. Pérez-Ruiz, L. N. L. de Lacalle, G. Urbikain, O. Pereira, S. Martínez, and J. Bris, "On the relationship between cutting forces and anisotropy features in the milling of LPBF Inconel 718 for near net shape parts," *Int J Mach Tools Manuf*, vol. 170, Nov. 2021, doi: 10.1016/j.ijmachtools.2021.103801.

[16]  A. Shrivastava, S. Anand Kumar, S. Rao, and B. K. Nagesha, "Exploring How LPBF process parameters impact the interface characteristics of LPBF Inconel 718 deposited on Inconel 718 wrought substrates," *Opt Laser Technol*, vol. 174, Jul. 2024, doi: 10.1016/j.optlastec.2024.110571.

[17]  Y. Zhao, K. Li, M. Gargani, and W. Xiong, "A comparative analysis of Inconel 718 made by additive manufacturing and suction casting: Microstructure evolution in homogenization," *Addit Manuf*, vol. 36, Dec. 2020, doi: 10.1016/j.addma.2020.101404.

[18]  J. Fu, S. Qu, J. Ding, X. Song, and M. W. Fu, "Comparison of the microstructure, mechanical properties and distortion of stainless steel 316 L fabricated by micro and conventional laser powder bed fusion," *Addit Manuf*, vol. 44, Aug. 2021, doi: 10.1016/j.addma.2021.102067.

[19]  M. R. Jandaghi, H. Pouraliakbar, S. H. Shim, V. Fallah, S. I. Hong, and M. Pavese, "In-situ alloying of stainless steel 316L by co-inoculation of Ti and Mn using LPBF additive manufacturing: Microstructural evolution and mechanical properties," *Materials Science and Engineering: A*, vol. 857, Nov. 2022, doi: 10.1016/j.msea.2022.144114.

[20]  P. He, C. Sun, and Y. Wang, "Material distortion in laser-based additive manufacturing of fuel cell component: Three-dimensional numerical analysis," *Addit Manuf*, vol. 46, Oct. 2021, doi: 10.1016/j.addma.2021.102188.

[21]  J. Gong, K. Wei, M. Liu, W. Song, X. Li, and X. Zeng, "Microstructure and mechanical properties of AlSi10Mg alloy built by laser powder bed fusion/direct energy deposition hybrid laser additive manufacturing," *Addit Manuf*, vol. 59, Nov. 2022, doi: 10.1016/j.addma.2022.103160.

[22]  K. Riener *et al.*, "Influence of particle size distribution and morphology on the properties of the powder feedstock as well as of AlSi10Mg parts produced by laser powder bed fusion (LPBF)," *Addit Manuf*, vol. 34, Aug. 2020, doi: 10.1016/j.addma.2020.101286.

[23] W. H. Kan, Y. Nadot, M. Foley, L. Ridosz, G. Proust, and J. M. Cairney, "Factors that affect the properties of additively-manufactured AlSi10Mg: Porosity versus microstructure," *Addit Manuf*, vol. 29, Oct. 2019, doi: 10.1016/j.addma.2019.100805.

[24] Q. Han *et al.*, "Additive manufacturing of high-strength crack-free Ni-based Hastelloy X superalloy," *Addit Manuf*, vol. 30, Dec. 2019, doi: 10.1016/j.addma.2019.100919.

[25] A. N. Jinoop *et al.*, "Influence of heat treatment on the microstructure evolution and elevated temperature mechanical properties of Hastelloy-X processed by laser directed energy deposition," *J Alloys Compd*, vol. 868, Jul. 2021, doi: 10.1016/j.jallcom.2021.159207.

[26] H. Wang *et al.*, "Micro-cracking, microstructure and mechanical properties of Hastelloy-X alloy printed by laser powder bed fusion: As-built, annealed and hot-isostatic pressed," *Addit Manuf*, vol. 39, Mar. 2021, doi: 10.1016/j.addma.2021.101853.

[27] A. Ghasemi, A. M. Kolagar, and M. Pouranvari, "Microstructure-performance relationships in gas tungsten arc welded Hastelloy X nickel-based superalloy," *Materials Science and Engineering: A*, vol. 793, Aug. 2020, doi: 10.1016/j.msea.2020.139861.

[28] J. Graneix, J. D. Beguin, J. Alexis, and T. Masri, "Weldability of Superalloys Hastelloy X by Yb: YAG Laser," *Adv Mat Res*, vol. 1099, pp. 61–70, Apr. 2015, doi: 10.4028/www.scientific.net/amr.1099.61.

[29] G. Gong *et al.*, "Comparative study on mechanical properties and high temperature oxidation behaviour of Hastelloy X and Inconel 718 fabricated by laser directed energy deposition," *Journal of Materials Research and Technology*, vol. 28, pp. 2624–2635, Jan. 2024, doi: 10.1016/j.jmrt.2023.12.199.

[30] A. Karimi, R. Soltani, M. Ghambari, and H. Fallahdoost, "High temperature oxidation resistance of plasma sprayed and surface treated YSZ coating on Hastelloy X," *Surf Coat Technol*, vol. 321, pp. 378–385, Jul. 2017, doi: 10.1016/j.surfcoat.2017.05.002.

[31] D. Chellaganesh, M. A. Khan, and J. T. W. Jappes, "Hot corrosion behaviour of nickel–iron-based superalloy in gas turbine application," *International Journal of Ambient Energy*, vol. 41, no. 8, pp. 901–905, Jul. 2020, doi: 10.1080/01430750.2018.1492446.

[32] G. Y. Lai, "96-TA-30 SEVERAL MODERN WROUGHT SUPERALLOYS FOR GAS TURBINE APPLICATIONS ABSTRACT," 1996. [Online]. Available: http://asmedigitalcollection.asme.org/GT/proceedings-pdf/TA1996/78774/V001T07A001/2568866/v001t07a001-96-ta-030.pdf

[33] Ł. Rakoczy, M. Grudzień, L. Tuz, K. Pańcikiewicz, and A. Zielińska-Lipiec, "Microstructure and Properties of a Repair Weld in a Nickel Based Superalloy Gas Turbine Component," *Advances in Materials Science*, vol. 17, no. 2, pp. 55–63, Jun. 2017, doi: 10.1515/adms-2017-0011.

[34] A. ali Malekbarmi, S. Zangeneh, and A. Roshani, "Assessment of premature failure in a first stage gas turbine nozzle," *Eng Fail Anal*, vol. 18, no. 5, pp. 1262–1271, Jul. 2011, doi: 10.1016/j.engfailanal.2011.03.011.

[35] A. M. Mirhosseini, S. Adib Nazari, A. Maghsoud Pour, S. Etemadi Haghighi, and M. Zareh, "Failure analysis of first stage nozzle in a heavy-duty gas turbine," *Eng Fail Anal*, vol. 109, Jan. 2020, doi: 10.1016/j.engfailanal.2019.104303.

[36] D. Yoon, I. Heo, J. Kim, S. Chang, and S. Chang, "Hold Time-Low Cycle Fatigue Behavior of Nickel Based Hastelloy X at Elevated Temperatures," *International Journal of Precision Engineering and Manufacturing*, vol. 20, no. 1, pp. 147–157, Jan. 2019, doi: 10.1007/s12541-019-00025-z.

[37] L. Lei, B. Li, H. Wang, G. Huang, and F. Xuan, "High-temperature high-cycle fatigue performance and machine learning-based fatigue life prediction of additively manufactured Hastelloy X," *Int J Fatigue*, vol. 178, Jan. 2024, doi: 10.1016/j.ijfatigue.2023.108012.

[38] H. U. Hong, I. S. Kim, B. G. Choi, H. W. Jeong, and C. Y. Jo, "Effects of temperature and strain range on fatigue cracking behavior in Hastelloy X," *Mater Lett*, vol. 62, no. 28, pp. 4351–4353, Nov. 2008, doi: 10.1016/j.matlet.2008.07.032.

[39] Y. L. Lu *et al.*, "Hold-time effects on low-cycle-fatigue behavior of hastelloy® X superalloy at high temperatures," in *Proceedings of the International Symposium on Superalloys*, Minerals, Metals and Materials Society, 2004, pp. 241–250. doi: 10.7449/2004/superalloys_2004_241_250.

[40] L. Zhao, X. Wang, L. Xu, Y. Han, and H. Jing, "Fatigue performance of Hastelloy X at elevated temperature via small punch fatigue test," *Theoretical and Applied Fracture Mechanics*, vol. 116, p. 103118, 2021.

[41] R. Huang, Y. Sun, C. Tan, D. Lin, X. Song, and H. Zhao, "Investigation of microstructure and failure mechanisms at room and elevated temperature of Hastelloy X produced by laser powder-bed fusion," *Next Materials*, vol. 2, p. 100142, Jan. 2024, doi: 10.1016/j.nxmate.2024.100142.

[42] H. Wang, B. Li, L. Lei, and F. Xuan, "Uncertainty-aware fatigue-life prediction of additively manufactured Hastelloy X superalloy using a physics-informed probabilistic neural network," *Reliab Eng Syst Saf*, vol. 243, Mar. 2024, doi: 10.1016/j.ress.2023.109852.

[43] J. Zhou, X. Han, H. Li, S. Liu, and J. Yi, "Investigation of layer-by-layer laser remelting to improve surface quality, microstructure, and mechanical properties of laser powder bed fused AlSi10Mg alloy," *Mater Des*, vol. 210, Nov. 2021, doi: 10.1016/j.matdes.2021.110092.

[44] H. M. Khan, Y. Karabulut, O. Kitay, Y. Kaynak, and I. S. Jawahir, "Influence of the post-processing operations on surface integrity of metal components produced by laser powder bed fusion additive manufacturing: a review," *Machining Science and Technology*, vol. 25, no. 1. Bellwether Publishing, Ltd., pp. 118–176, 2020. doi: 10.1080/10910344.2020.1855649.

[45] Y. Yin, J. Zhang, J. Gao, Z. Zhang, Q. Han, and Z. Zan, "Laser powder bed fusion of Ni-based Hastelloy X superalloy: Microstructure, anisotropic mechanical properties and strengthening mechanisms," *Materials Science and Engineering: A*, vol. 827, Oct. 2021, doi: 10.1016/j.msea.2021.142076.

[46] D. Jiang, Y. Tian, Y. Zhu, and A. Huang, "Investigation of surface roughness post-processing of additively manufactured nickel-based superalloy Hastelloy X using electropolishing," *Surf Coat Technol*, vol. 441, Jul. 2022, doi: 10.1016/j.surfcoat.2022.128529.

[47] Y. Tian, D. Tomus, P. Rometsch, and X. Wu, "Influences of processing parameters on surface roughness of Hastelloy X produced by selective laser melting," *Addit Manuf*, vol. 13, pp. 103–112, Jan. 2017, doi: 10.1016/j.addma.2016.10.010.

[48] R. Esmaeilizadeh *et al.*, "On the effect of laser powder-bed fusion process parameters on quasi-static and fatigue behaviour of Hastelloy X: A microstructure/defect interaction study," *Addit Manuf*, vol. 38, Feb. 2021, doi: 10.1016/j.addma.2020.101805.

[49] E. Maleki *et al.*, "Fatigue behaviour of notched laser powder bed fusion AlSi10Mg after thermal and mechanical surface post-processing," *Materials Science and Engineering: A*, vol. 829, Jan. 2022, doi: 10.1016/j.msea.2021.142145.

[50] A. du Plessis and S. Beretta, "Killer notches: The effect of as-built surface roughness on fatigue failure in AlSi10Mg produced by laser powder bed fusion," *Addit Manuf*, vol. 35, Oct. 2020, doi: 10.1016/j.addma.2020.101424.

[51] O. Scott-Emuakpor, T. George, E. Henry, C. Holycross, and J. Brown, "As-Built Geometry and Surface Finish Effects on Fatigue and Tensile Properties of Laser Fused Titanium 6Al-4V." [Online]. Available: http://asmedigitalcollection.asme.org/GT/proceedings-pdf/GT2017/50916/V006T24A002/2434270/v006t24a002-gt2017-63482.pdf

[52] M. H. Raymond, L. F. Coffin, and M. Asme, "Geometrical Effects in Strain Cycled Aluminum," 1963. [Online]. Available: http://asmedigitalcollection.asme.org/fluidsengineering/article-pdf/85/4/548/5763410/548_1.pdf

[53] P. S. Maiya, D. E. Busch, D. E. Busch Are Metallurgist, and S. Techni-Cian, "Effect of Surface Fatigue Behavior Roughness on Low-Cycle of Type 304 Stainless Steel."

[54] C. Elangeswaran *et al.*, "Predicting fatigue life of metal LPBF components by combining a large fatigue database for different sample conditions with novel simulation strategies," *Addit Manuf*, vol. 50, Feb. 2022, doi: 10.1016/j.addma.2021.102570.

[55] E. Maleki *et al.*, "On the efficiency of machine learning for fatigue assessment of post-processed additively manufactured AlSi10Mg," *Int J Fatigue*, vol. 160, Jul. 2022, doi: 10.1016/j.ijfatigue.2022.106841.

[56] N. Macallister and T. H. Becker, "Fatigue life estimation of additively manufactured Ti-6Al-4V: Sensitivity, scatter and defect description in Damage-tolerant models," *Acta Mater*, vol. 237, Sep. 2022, doi: 10.1016/j.actamat.2022.118189.

[57] "Standard Test Method for Strain-Controlled Fatigue Testing 1." doi: 10.1520/E0606_E0606M-21.

[58] U. Guide, T. Ferreira, and W. Rasband, "ImageJ 1.44 The ImageJ User Guide 1.44." 2011. [Online]. Available: http://imagej.nih.gov/ij/docs/guide/

[59] "Avizo Basic User Guide."

[60] E. S. Gadelmawla, M. M. Koura, T. M. A. Maksoud, I. M. Elewa, and H. H. Soliman, "Roughness parameters."

[61]  K. Cooray, "Generalized gumbel distribution," *J Appl Stat*, vol. 37, no. 1, pp. 171–179, Jan. 2010, doi: 10.1080/02664760802698995.

[62]  H. Suzuki, T. Iseki, and Y. Shoda, "High-temperature low-cycle fatigue tests on hastelloy x," *J Nucl Sci Technol*, vol. 14, no. 5, pp. 381–386, 1977, doi: 10.1080/18811248.1977.9730773.

[63]  H. K. Wiest and S. D. Heister, "Experimental study of gas turbine combustion with elevated fuel temperatures," *J Eng Gas Turbine Power*, vol. 136, no. 12, 2014, doi: 10.1115/1.4027907.

[64]  J. D. G. Sumpter, "Pop-in fracture: observations on load drop, displacement increase, and crack advance," *Int J Fract*, vol. 49, pp. 203–224, 1991.

[65]  A. T. RJH Wanhill Hattenberg, "Executive summary," 2006. [Online]. Available: www.nlr.nl

[66]  R. G. Forman *et al.*, "Growth Behavior of Surface Cracks in the Circumferential Plane of Solid and Hollow Cylinders 'Growth Behavior of Surface Cracks in the Cireunfferentlai Plane of Solid and Hollow Cylinders,' Fracture Mechan-ics: Seventeenth Volume, ASTM STP 905," 1986. [Online]. Available: www.astm.org

[67]  A. N. Jinoop, J. Denny, C. P. Paul, J. Ganesh Kumar, and K. S. Bindra, "Effect of post heat-treatment on the microstructure and mechanical properties of Hastelloy-X structures manufactured by laser based Directed Energy Deposition," *J Alloys Compd*, vol. 797, pp. 399–412, Aug. 2019, doi: 10.1016/j.jallcom.2019.05.050.

[68]  L. Hitzler, C. Janousch, J. Schanz, M. Merkel, F. Mack, and A. Öchsner, "Non-destructive evaluation of AlSi10Mg prismatic samples generated by selective laser melting: Influence of manufacturing conditions," in *Materialwissenschaft und Werkstofftechnik*, Wiley-VCH Verlag, Jun. 2016, pp. 564–581. doi: 10.1002/mawe.201600532.

[69]  S. Romano *et al.*, "Fatigue strength estimation of net-shape L-PBF Co–Cr–Mo alloy via non-destructive surface measurements," *Int J Fatigue*, vol. 178, Jan. 2024, doi: 10.1016/j.ijfatigue.2023.108018.

[70]  H. Chen *et al.*, "Correlation model between surface defects and fatigue behavior of 2024 aluminum alloy," *Int J Fatigue*, vol. 168, Mar. 2023, doi: 10.1016/j.ijfatigue.2022.107379.

[71]  B. Vieille, A. Duchaussoy, S. Benmabrouk, R. Henry, and C. Keller, "Fracture behavior of Hastelloy X elaborated by laser powder bed fusion: Influence of microstructure and building direction," *J Alloys Compd*, vol. 918, Oct. 2022, doi: 10.1016/j.jallcom.2022.165570.

[72]  N. W. Sachs, "Understanding the surface features of fatigue fractures: How they describe the failure cause and the failure history," *Journal of Failure Analysis and Prevention*, vol. 5, no. 2. pp. 11–15, Apr. 2005. doi: 10.1361/15477020522924.

[73]  J. Yue, Y. Dong, and C. Guedes Soares, "An experimental-finite element method based on beach marks to determine fatigue crack growth rate in thick plates with varying stress states," *Eng Fract Mech*, vol. 196, pp. 123–141, Jun. 2018, doi: 10.1016/j.engfracmech.2018.04.015.

[74]  M. D. Sangid, "The physics of fatigue crack initiation," *Int J Fatigue*, vol. 57, pp. 58–72, 2013, doi: 10.1016/j.ijfatigue.2012.10.009.

[75]   E. Bitzek and P. Gumbsch, "Mechanisms of dislocation multiplication at crack tips," *Acta Mater*, vol. 61, no. 4, pp. 1394–1403, Feb. 2013, doi: 10.1016/j.actamat.2012.11.016.

[76]   R. Esmaeilizadeh *et al.*, "Fatigue Characterization and Modeling of Additively Manufactured Hastelloy-X Superalloy," *J Mater Eng Perform*, vol. 31, no. 8, pp. 6234–6245, Aug. 2022, doi: 10.1007/s11665-022-06595-w.

[77]   Z. Guo, N. Saunders, P. Miodownik, and J.-P. Schille, "Proceedings of CREEP8 Eighth International Conference on Creep and Fatigue at Elevated Temperatures DRAFT(Revised) PVP200726074 MODELLING THE STRAIN-LIFE RELATIONSHIP OF COMMERCIAL ALLOYS."